\documentclass[pre,twocolumn,superscriptaddress,citeautoscript]{revtex4-1}

\usepackage{epsfig}
\usepackage[usenames, dvipsnames]{color}
\usepackage{natbib}
\usepackage{xifthen}
\usepackage{bm}
\usepackage[version=3]{mhchem} 
\usepackage{mciteplus}
\usepackage{multirow}
\usepackage{graphicx}
\usepackage{dcolumn}
\usepackage{bm}
\usepackage{mathptmx,amsmath,amssymb,txfonts}
\usepackage{subfig}
\usepackage{blindtext}
\begin{document}
\title{{From Sticky-Hard-Sphere to Lennard-Jones-Type Clusters}}
\date{\today}

\author{Lukas Trombach}
\affiliation{Centre for Theoretical Chemistry and Physics, The New Zealand Institute for Advanced Study, Massey University Auckland, Private Bag 102904, 0632 Auckland, New Zealand}
\author{Robert S. Hoy}
\affiliation{Department of Physics, University of South Florida, Tampa, Florida 33620, USA}
\author{David J. Wales}
\affiliation{University Chemical Laboratories, Lensfield Road, Cambridge CB2 1EW, UK}
\author{Peter Schwerdtfeger}
\email[Email: ]{p.a.schwerdtfeger@massey.ac.nz}
\affiliation{Centre for Theoretical Chemistry and Physics, The New Zealand Institute for Advanced Study, Massey University Auckland, Private Bag 102904, 0632 Auckland, New Zealand}
\affiliation{Centre for Advanced Study (CAS) at the Norwegian Academy of Science and Letters, Drammensveien 78, NO-0271 Oslo, Norway}

\begin{abstract}
A relation $\mathcal{M}_{\mathrm{SHS}\to\mathrm{LJ}}$ between the set of non-isomorphic sticky hard sphere clusters
    $\mathcal{M}_\mathrm{SHS}$ and the sets of local energy minima $\mathcal{M}_{LJ}$ of the $(m,n)$-Lennard-Jones potential 
    $V^\mathrm{LJ}_{mn}(r) = \frac{\varepsilon}{n-m} [ m r^{-n} - n r^{-m} ]$
    is established. The number of nonisomorphic stable clusters depends strongly and nontrivially on both $m$ and $n$,
    and increases exponentially with increasing cluster size $N$ for $N \gtrsim 10$. 
    While the map from $\mathcal{M}_\mathrm{SHS}\to
    \mathcal{M}_{\mathrm{SHS}\to\mathrm{LJ}}$ is non-injective and non-surjective, the number of
    Lennard-Jones structures missing from the map is relatively small for
    cluster sizes up to $N=13$, and most of the missing structures correspond
    to energetically unfavourable minima even for fairly low $(m,n)$.  
    Furthermore, even the softest Lennard-Jones potential predicts that the coordination of 13 spheres
    around a central sphere is problematic (the Gregory-Newton problem). 
    A more realistic extended Lennard-Jones potential chosen from
coupled-cluster calculations for a rare gas dimer 
    leads to a substantial increase in the number of nonisomorphic clusters, even though the potential curve is very similar 
    to a (6,12)-Lennard-Jones potential. 
\end{abstract}
\maketitle

\section{Introduction}

The nucleation of atoms and molecules in the gas phase, or liquid, to the solid
state is still an active research field
\cite{Stillinger-1984,Martin-1996,Wales-1996,Vlieg-2007,Arkus-2010,Woodley-2010,Karthika-2016,holmes17}.
Rowland noted in 1949 that ``The gap between theory and the experimental
approaches to nucleation has been too wide'' and ``the subject [nucleation] is
still in the alchemical stage'' \cite{Rowland-1949}. More than half a century
later, despite all the advancements made in cluster physics, 
``there is still a large gap between experiment
and theory'' as Unwin noted \cite{Unwin-2007}.

The underlying reason for this rather slow progress is that cluster formation
is a dynamic process, and fully characterizing the corresponding
high-dimensional potential energy landscape is typically an NP-hard problem,
since there are (presumably) exponentially many local minima at any given
temperature and pressure
\cite{Stillinger-1984,Oganov-2006,Massen-2007,wales10,Oganov-2011,calvo12,Wales-2015}.
Moreover, phase transitions between different morphologies as a function of
size $N$ usually occur where $N$ is too large for an accurate
quantum-theoretical treatment
\cite{waal89,clevelandl91,vandewaal96a,Doye-1995,vandeWaalTd00}.  For example,
Krainyukova experimentally studied the growth of argon clusters
\cite{Krainyukova-2012}, and found that small, initially icosahedral clusters
transform into anti-Mackay clusters for $N>2000$, and finally into the closed
packed fcc or hcp structures at $N>10^5$ atoms, in qualitative agreement with
theoretical predictions using Lennard-Jones (LJ) type potentials
\cite{Martin-1996,Schwerdtfeger-2006,Krainyukova-2007}.  The notorious {\it
rare gas problem} was solved only very recently by accurate relativistic
quantum methods, correctly predicting a slight preference of the fcc over the
hcp phase due to phonon dispersion \cite{Schwerdtfeger-2016}.

Simple models often have to be used to simulate cluster growth and nucleation \cite{Johnston-1999,Shibuta-2015,Leitold-2016,Sweatman-2016}.
The simplest model potentials that can be applied to theoretical studies of
atomic cluster formation are ``HCR-SRA'' potentials with isotropic
hard-core-like repulsive and short-range-attractive interactions
\cite{baxter68}.  The simplest HCR-SRA potential is the ``sticky hard sphere''
(SHS) potential \cite{yuste93}
\begin{align}
    V_\mathrm{SHS}(r)=\begin{cases}
        \infty, & r < r_s,\\
        -\varepsilon, & r = r_s,\\
        0, & r > r_s,
    \end{cases}
\label{eqn:KS}
\end{align}
where $r_s$ and $\varepsilon$ can be arbitrarily set to 1 (unit sphere and reduced
units, respectively).  Eq.\ \ref{eqn:KS} can be used as a perturbative basis
for finite-ranged HCR-SRA potentials \cite{cochran06,holmes13}.  Since sticky
hard spheres are impenetrable and their energy $E = -N_c\varepsilon$ ~is a
function only of the number of interparticle contacts $N_c$, SHS cluster
structure and energetics can be uniquely mapped to their adjacency
matrices $\bar{A}$, where $N_c=\sum_{i<j}^N A_{ij}$.
This mapping allows them to be exactly characterized via complete enumeration
\cite{arkus_minimal_2009,arkus_deriving_2011,hoy_structure_2012}; recent
studies have identified all mechanically stable SHS clusters for $N \leq 14$,
and putatively complete sets for $N \leq 19$
\cite{arkus_minimal_2009,arkus_deriving_2011,hoy_structure_2012,hoy15,holmes16,kallus17}.
Note, however, that different SHS structures can have the same adjacency
matrix for $N \geq 14$ \cite{holmes16}, and the mapping is therefore only surjective.

From the Gregory-Newton kissing-number argument proved in 1953 by Sch\"utte and
van der Waerden \cite{Schutte-1952}, no sphere can be surrounded by more than
12 spheres of equal radius \cite{conway-2013book}.  For small clusters,
graph-theoretic arguments dictate $\mathrm{max}(N_c)\le N(N-1)/2$.  Thus a
loose bound on the maximum contact number $N_c(N)$ is
\begin{equation}
    N_c^\mathrm{max}(N) \le \mathrm{min}\{N(N-1)/2,f(N)\}
    \label{eqn:upperlimitNc}
\end{equation}
with $f(N)=6N$.  This upper bound has been tightened several times, most
recently by Bezdek and Reid \cite{Bezdek-2013} to
\begin{equation}
    f(N)=6N-3(18)^{1/3}\pi^{-2/3}N^{2/3}.
    \label{eqn:upperlimitBR}
\end{equation}
In
Refs.~\cite{hoy15,holmes16}
it was shown that $N_c^\mathrm{max}(N) =
\{6,9,12,15,18,21,25,29,33,36,40,44,48,52,56,60\}$ for $4 \leq N \leq 19$.
While determining $N_c^\mathrm{max}(N)$ for arbitrary $N$ is equivalent to the
still-unsolved Erd\"os unit distance problem \cite{Erdos-1946}, it is clear
that $N_c^\mathrm{max}(N) = 3N - 6 + m(N)$, where $m(N)$ grows slowly from zero
to around $f(N) - (3N-6)$ with increasing $N$.

While the maximum contact number increases (sub)linearly with $N$, the number of non-isomorphic cluster structures $|\mathcal{M}(N)|$ and 
transition states is assumed to increase exponentially \cite{Stillinger-1999,Oganov-2006,Forman-2017} (here we denote $\mathcal{M}(N)$ as the set of all non-isomorphic cluster structures of size $N$, and 
$|\mathcal{M}(N)|$ as the number of structures in $\mathcal{M}(N)$).
Stillinger showed that under certain conditions $\lim_{N\to\infty} |\mathcal{M}(N)| \propto \exp(\alpha N)$ \cite{Stillinger-1999}.
For SHS clusters, the complete set $\mathcal{M}_\mathrm{SHS}(N,N_c)$ has been exactly
determined for $N \leq 14$ and $3N - 6 \leq N_c \leq N_c^\mathrm{max}(N)$ via exact
enumeration studies employing geometric rejection rules \cite{hoy15,holmes16}.
Unfortunately, such precise calculations are very difficult for finite-ranged
potentials since exhaustive searches for energy minima 
are computationally intensive \cite{heiles_global_2013}.  
Only a few such studies have been performed, e.g.\ recent studies of $N \leq 19$ clusters interacting via short-range Morse
potentials \cite{wales10,calvo12,C7CP03346J}.

\begin{figure}
    \includegraphics[width=\columnwidth]{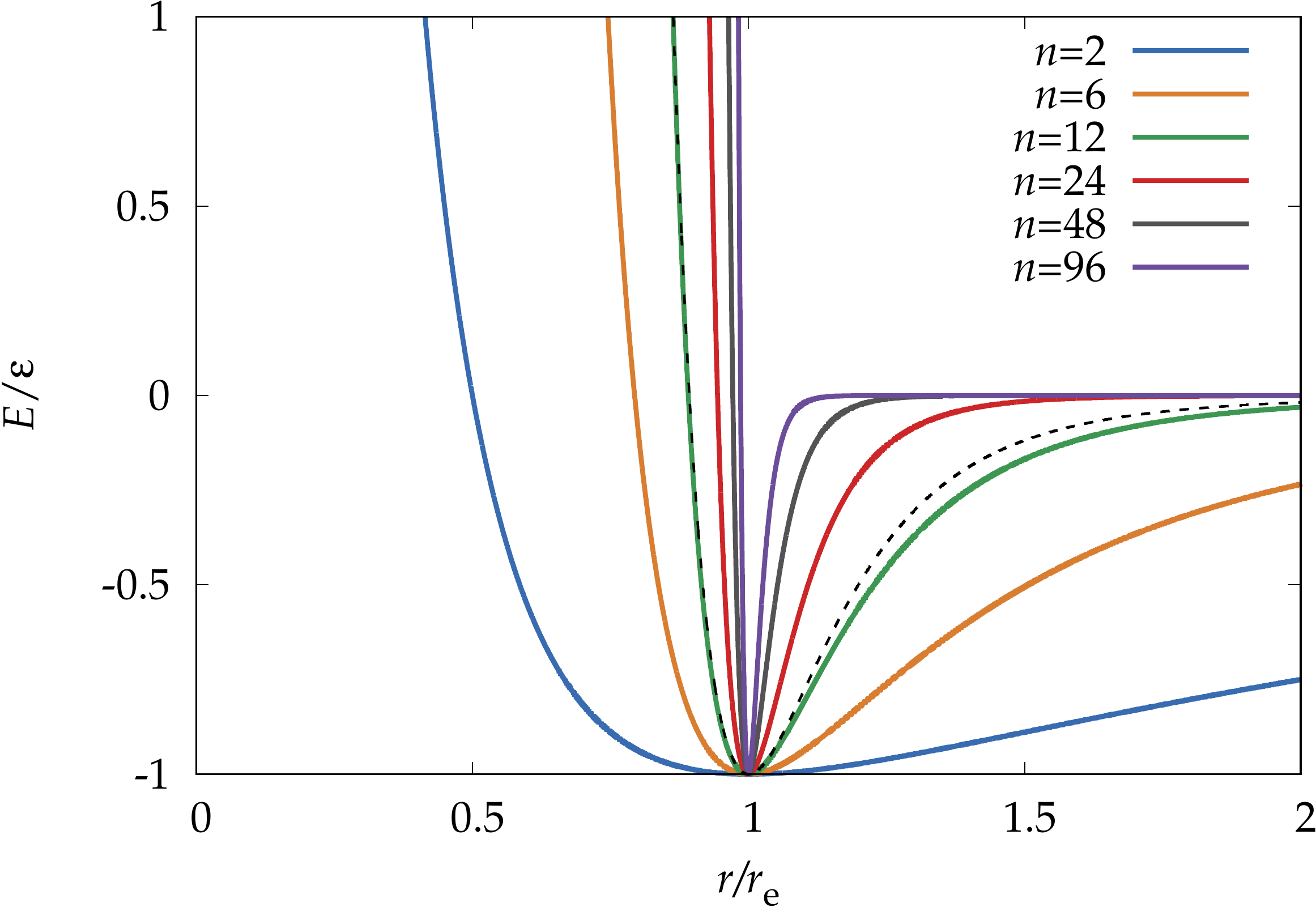}
    \caption{Lennard-Jones potentials for different exponents $(m,n)$ with
    fixed $n=2m$. As the exponents grow larger the well of attraction becomes narrower 
    and its shape approaches the SHS potential. The dashed line
    shows the extended LJ potential for the xenon dimer \cite{jerabek_relativistic_2017}.}
    \label{fig:LJ}
\end{figure}

It remains unclear how the HCR-SRA models commonly
used in cluster physics relate to more physically relevant, softer interaction potentials such
as the $(m,n)$-Lennard-Jones (LJ) form:
\begin{equation}
V_{m,n}^\mathrm{LJ}(r)=\frac{\varepsilon}{n-m}\left[m\left(\frac{r_e}{r}\right)^{n}-n\left(\frac{r_e}{r}\right)^{m}\right] \ \ \ \ \ \ \ \ \ \  ({\rm with}\ n > m).
\label{eqn:nmpot}
\end{equation}
Here $\varepsilon>0$ is the dissociation energy and $r_e$ the equilibrium
two-body interparticle distance. To simplify the presentation,
we (without loss of generality) adopt reduced units ($\varepsilon=1$, $r_e=1$) below.
For $m,n\rightarrow \infty$,
$V_{m,n}^\mathrm{LJ}(r) \rightarrow V_\mathrm{SHS}(r)$ (Fig.~\ref{fig:LJ}); the
energy landscapes of the two potentials converge in this limit.  However, real
systems are not in this limit.  For example, for $N = 13$, there are $|\mathcal{M}_\mathrm{SHS}|=97,221$
stable SHS clusters \cite{hoy15,holmes16},
but only $|\mathcal{M}_\mathrm{LJ}|=1,510$ stable $(m,n) = (6,12)$ LJ clusters \cite{doye_evolution_1999}.  
This difference is understood qualitatively -- energy landscapes are well known to support more
local minima
as the range of the interaction potential decreases \cite{braier90,Wales01}.
There are several effects that will cause the set of
stable LJ clusters to increasingly deviate from the set of stable SHS clusters
as interactions become longer ranged.  As $n$ and $m$ decrease,
second-nearest-neighbor attractions become increasingly important,
producing stable structures with $r_{ij} \leq 1$.  Fold catastrophes
\cite{Wales01,wales04} progressively eliminate stable SHS clusters, and several stable SHS
structures may collapse into a single stable LJ cluster.  However,
detailed quantitative understanding of such effects remains rather limited.

In this paper, we quantitatively examine how stable $N \leq 14$ LJ cluster
structures evolve away from the SHS limit as ($m, n$) decrease.  We focus on
both the topography of the energy landscape (decreasing
$|\mathcal{M}_{\rm LJ}(N)|$) and the evolving topologies of the stable cluster sets.
We examine these changes in further detail for specific $N = 13-14$ clusters discussed by
Gregory and Newton in the 1600s in the context of the kissing number problem
\cite{Schutte-1952}, and also for a more realistic two-body potential that has
been shown to accurately model rare-gas clusters \cite{Schwerdtfeger-2006}.

\section{Computational Methods}

The \textit{pele} program~\cite{_pele:_2017} was used to generate putatively
complete sets of local minima for $(m,n)$-Lennard-Jones potentials 
$V_{mn}^{\rm LJ}(r)$ as defined in Eq.(\ref{eqn:nmpot}).
This program  applies a basin-hopping algorithm that divides the potential energy
surface into basins of attraction, effectively mapping each point in
configuration space to a local minimum structure
\cite{lis87,waless99,wales_global_1997}.  The results confirmed the
number of local minima reported in previous work \cite{doye_saddle_2002}.
Finite computer time limited our search to clusters of size $N \leq 13$.

Starting from the sticky hard sphere packings up to $N=14$, with Cartesian
coordinates given by the exact enumeration algorithm \cite{hoy_structure_2012}
including rigid hypostatic clusters ($N_c<3N-6$) \cite{holmes16}, we carried out geometry
optimisations with $(m,n)$-Lennard-Jones potentials using the multidimensional function minimiser from the
C++ library \textit{dlib} \cite{king_dlib-ml:_2009}. The optimisation scheme
was either the Broyden-Fletcher-Goldfarb-Shanno (BFGS) or the conjugate
gradient (CG) algorithm.  The optimisations were terminated when the change in
energy (in reduced units) over the course of one optimization cycle was smaller
than $10^{-15}$. 
Subsequently, the eigenvalues of the Hessian were checked
for all stationary points. If negative eigenvalues were found, the affected
structures were reoptimized following displacements in both directions along
the corresponding eigenvectors to locate true local minima. This procedure assures
that the floppy SHS packings are successfully mapped into LJ minima.

As the optimisations often result in many duplicates, especially for small
values of $n$ and $m$ where we have
$|\mathcal{M}_{(m,n)\mathrm{-LJ}}| \ll |\mathcal{M}_\mathrm{SHS}|$, the final
structures were further analysed and sorted. Nonisomorphic SHS clusters can be
distinguished (apart from permutation of the particles) by their different
adjacency matrices for $N \leq 13$ \cite{holmes16}.  This is not the case for
soft potentials like the LJ potential since drawing edges (bonds) between the
vertices (atoms) becomes a matter of defining the distance cutoff criterion
for a bond to be drawn. Therefore, we compare the interparticle distances $\{r_{ij}\}$
instead: two clusters are isomorphic (structurally identical) if they have the same ordered
set of inter-particle distances $\{r_{ij}\}$.  While enantiomers cannot be
separated using this methodology, permutation-inversion isomers are usually
lumped together, since the number of distinct minima is analytically related to
the order of the corresponding point group \cite{wales04}.  To verify the
number of distinct structures we introduced a second ordering scheme using the
energy and moment of inertia tensor eigenvalues. 

Two sets of structures are obtained from our optimization procedure: the first
set contains all possible LJ minima $\mathcal{M}_\mathrm{LJ}$ from the basin-hopping
algorithm, while the second set $\mathcal{M}_\mathrm{SHS\to LJ}$ contains the
LJ minima obtained using only the $\mathcal{M}_\mathrm{SHS}$ sticky-hard-sphere cluster structures as starting points for the geometry optimization.
To compare and identify corresponding structures between the two sets, the
$N(N-1)/2$ inter-particle distances $\{r_{ij}\}$ were again used as an identifying fingerprint.

Two-body ``extended Lennard-Jones'' (ELJ) potentials that accurately model two-body interactions in rare-gas clusters can be written as expansions of inverse-power-law terms \cite{Schwerdtfeger-2006}:
\begin{equation} \label{eq:ELJ}
V_{\rm ELJ}(r)=\sum_{n} c_nr^{-n},
\end{equation}
where in reduced units the condition $\sum_{n} c_n=-1$ holds.
For comparison to the simple (6,12)-LJ potential, we used the ELJ potential
derived from relativistic coupled-cluster theory applied to the xenon dimer,
with the following coefficients for the ELJ potential (in reduced units):
$c_6=-1.0760222355$; $c_8=-1.4078314494$; $c_9=-185.6149933139$;
$c_{10}=+1951.8264493941$; $c_{11}=-8734.2286559729$;
$c_{12}=+22273.3203327203$; $c_{13}=-35826.8689874832$;
$c_{14}=+37676.9744744424$; $c_{15}=-25859.2842295062$;
$c_{16}=+11157.4331408911$; $c_{17}=-2745.9740079192$; $c_{18}=+293.9003309498$
\cite{jerabek_relativistic_2017}. The ELJ potential for xenon is shown in
Figure \ref{fig:LJ} (dashed line).

\section{Results}

\subsection{Exploring the limits of Lennard-Jones}

\begin{figure}
    \centering
    \subfloat{\includegraphics[width=\columnwidth]{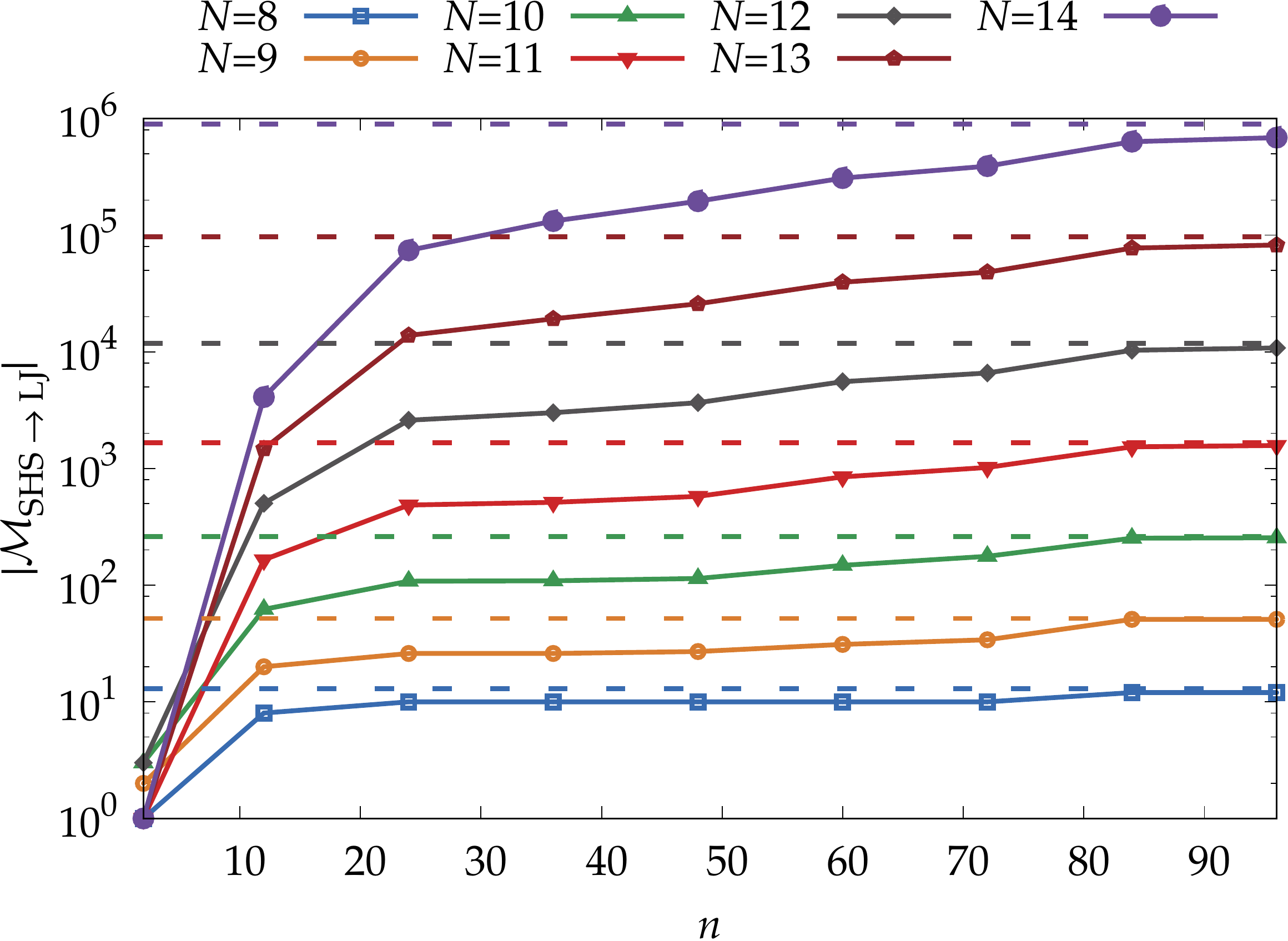}}\\
    \subfloat{\includegraphics[width=\columnwidth]{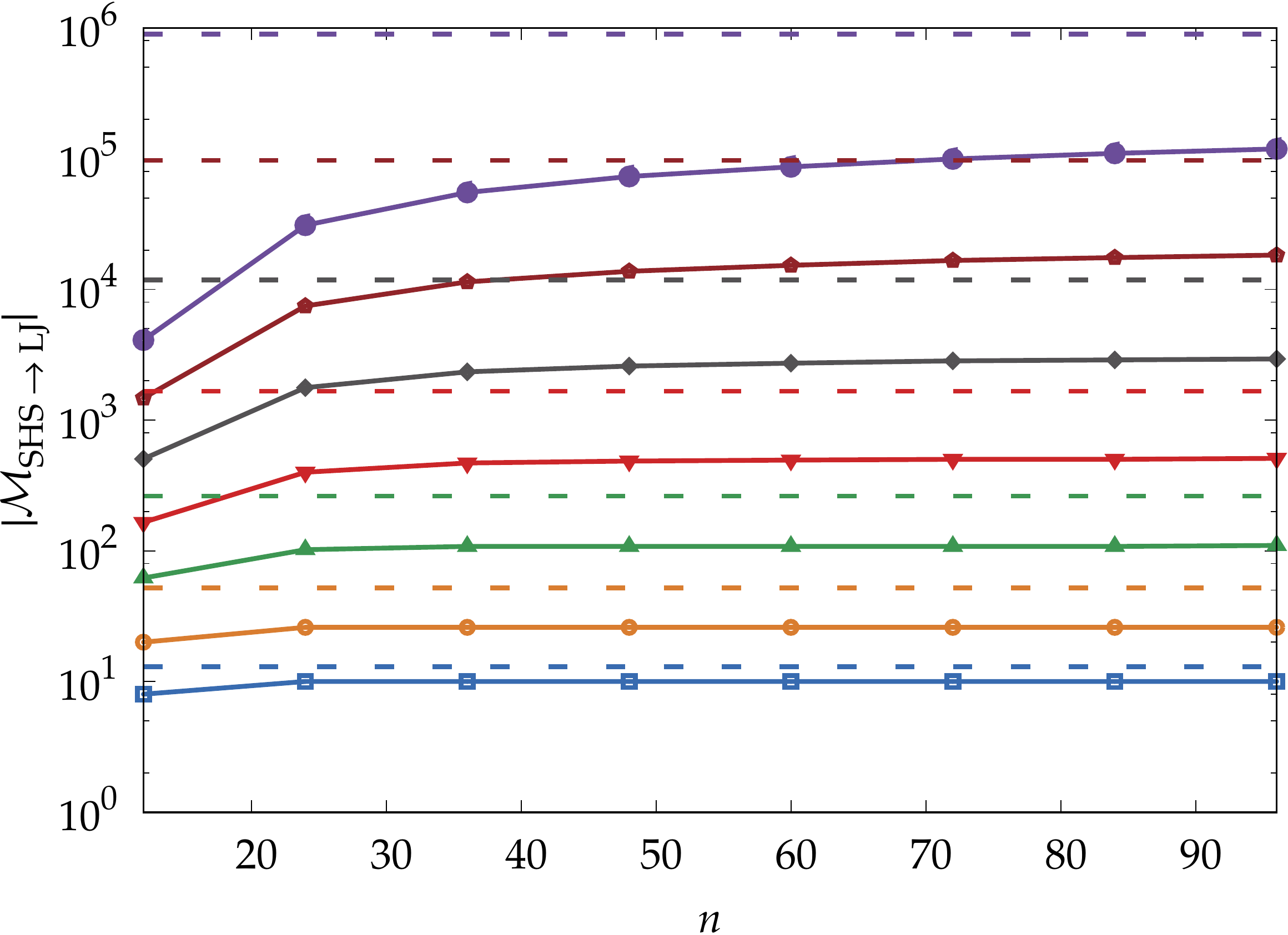}}
    \caption{
    Convergence of the number of distinct LJ local minima
 $|\mathcal{M}_\mathrm{SHS\to LJ}|$
     obtained through geometry optimisations starting from the nonisomorphic SHS structures with
     increasing LJ exponent $n$. 
Permutation-inversion isomers and enantiomers are not distinguished.     
     The dashed line gives the exact SHS limit $|\mathcal{M}_\mathrm{SHS}|$. 
     Top panel: $m=n/2$. Bottom panel: fixed $m=6$.}
    \label{fig:expinfty}
\end{figure}

To study the convergence behavior of the number of distinct (nonisomorphic) LJ minima in
the SHS limit, we performed geometry optimisations,
starting from all nonisomorphic SHS structures.  We will show later that the
number of unique minima obtained in this procedure $|\mathcal{M}_\mathrm{SHS\to
LJ}|$ only misses out on a small portion of minima obtained from the more exhaustive
basin-hopping approach, i.e.~$|\mathcal{M}_\mathrm{SHS\to
LJ}|\approx|\mathcal{M}_\mathrm{LJ}|$.  The results for a constant chosen ratio
of LJ exponents $n/m=2$ are shown in Figure~\ref{fig:expinfty} (top). 

$|\mathcal{M}_\mathrm{SHS\to LJ}|$ smoothly  converges towards the SHS limit (dashed line, values in Table \ref{tab:comp}) from below, 
thus demonstrating that for LJ systems the number of distinct minima does not grow faster than exponentially.  
The (48,96)-LJ potential has $\Delta\mathcal{M} \equiv |\mathcal{M}_\mathrm{LJ}| - |\mathcal{M}_\mathrm{SHS\to LJ}| = \{1,1,7,91,1019,14890,209938\}$ fewer stable minima than the SHS potential.
The fractions of missing minima $\Delta\mathcal{M}/|\mathcal{M}_\mathrm{SHS}|$ for this potential grow with increasing $N$ and are respectively $\{7.69,1.92,2.67,5.46,8.62,15.32,23.44\}\%$.  Note that for $N \geq 10$ most of these missing minima correspond to high energy ($N_c < N_c^\mathrm{max}$) structures.

If the exponent $n$ for the repulsive part of the LJ potential is increased
with $m$ kept constant, the LJ potential becomes equivalent to the SHS
potential in the repulsive range but remains attractive at long range.
Figure~\ref{fig:expinfty} (bottom) shows the convergence of the number of
unique structures with respect to $n$ at set $m=6$ towards the SHS limit. Here, the
number of distinct minima converges towards a number that is much smaller
than the total number of SHS packings demonstrating that (as expected) the
attractive part of the potential contributes significantly to the decrease of
the number of local minima compared to the rigid SHS model.

\begin{figure}
    \centering
    \includegraphics[width=\columnwidth]{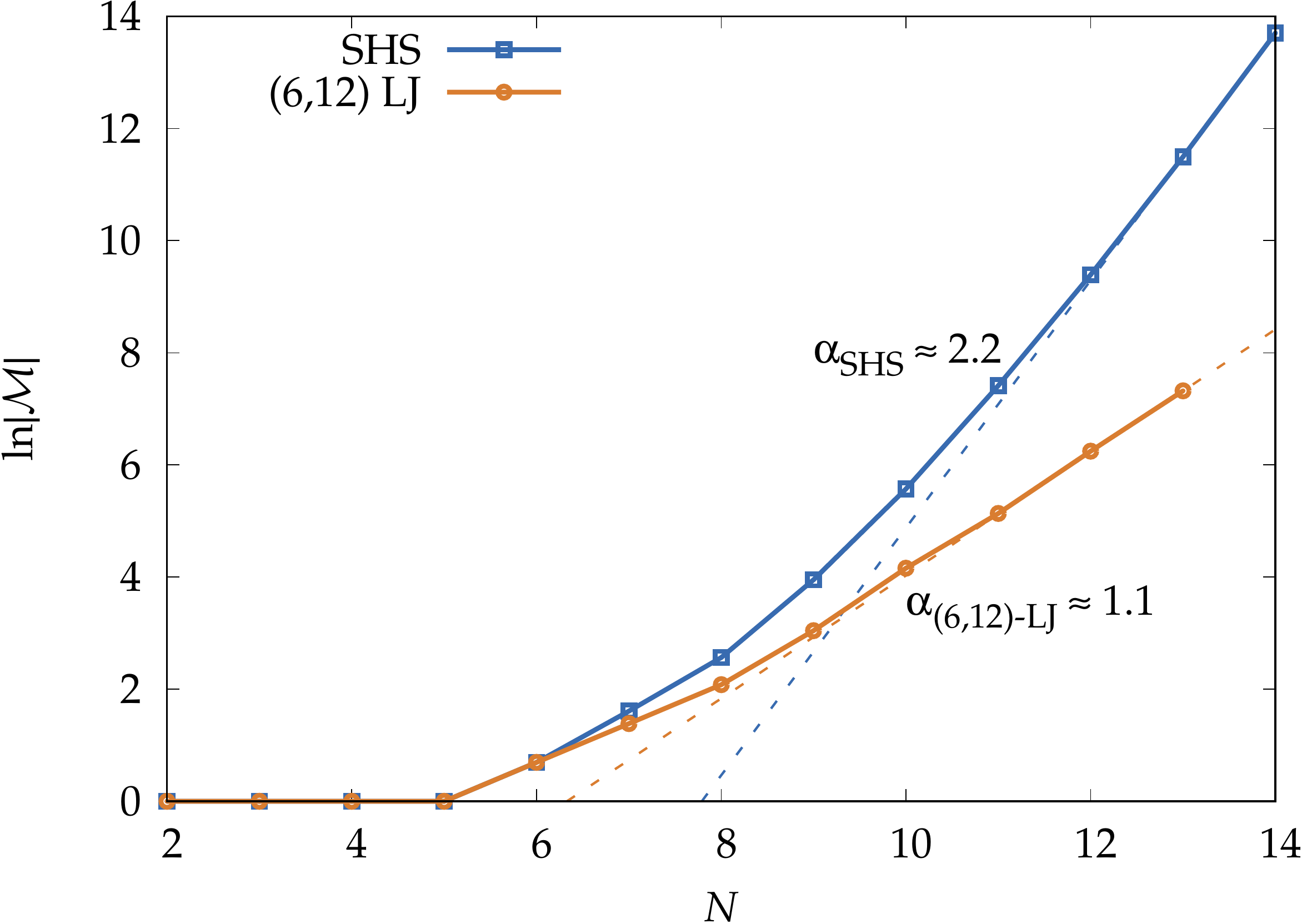}
    \caption{Growth behaviour of $|\mathcal{M}(N)|$ of SHS and (6,12)-LJ clusters and 
    corresponding asymptotic exponential rise rate parameter $\alpha$ for $N \geq 12$ as defined in Eq.(\ref{eq:Stil}).
    The intercepts ln$|\mathcal{M}(N=0)|$ are $-17.19$ and $-6.94$ for the SHS and (6,12)-LJ cases respectively.}
    \label{fig:asympt}
\end{figure}

To see if the asymptotic increase in the number of distinct minima $|\mathcal{M}(N)| \sim e^{\alpha N}$ 
is indeed exponential, we use Stillinger's expression for the asymptotic exponential rise rate parameter \cite{Stillinger-1999}
\begin{equation} \label{eq:Stil}
\alpha = \lim_{N\rightarrow \infty} \left( N^{-1} \mathrm{ln} |\mathcal{M}(N)| \right).
\end{equation}
Figure \ref{fig:asympt} shows the number of distinct minima for SHS clusters
obtained from the data shown in Table \ref{tab:comp}.  The $N \geq 12$ SHS data
gives $\alpha_\mathrm{SHS}\approx 2.21$. Figure \ref{fig:asympt} also shows the
(6,12)-LJ results obtained using basin-hopping; these yield
$\alpha_\mathrm{LJ}\approx 1.10$, which is close to the $\alpha=0.8$ value
estimated by Wallace \cite{Wallace-1997} or to the recently given value of 1.04
by Forman and Cameron \cite{Forman-2017}.  Note that the rapid increase of
$|\mathcal{M}_\mathrm{SHS}|/|\mathcal{M}_\mathrm{LJ}|$ with $N$ is explained by
the much larger values of $\alpha$ for the SHS compared to the LJ clusters.

\begin{figure}
    \centering
    \includegraphics[width=\columnwidth]{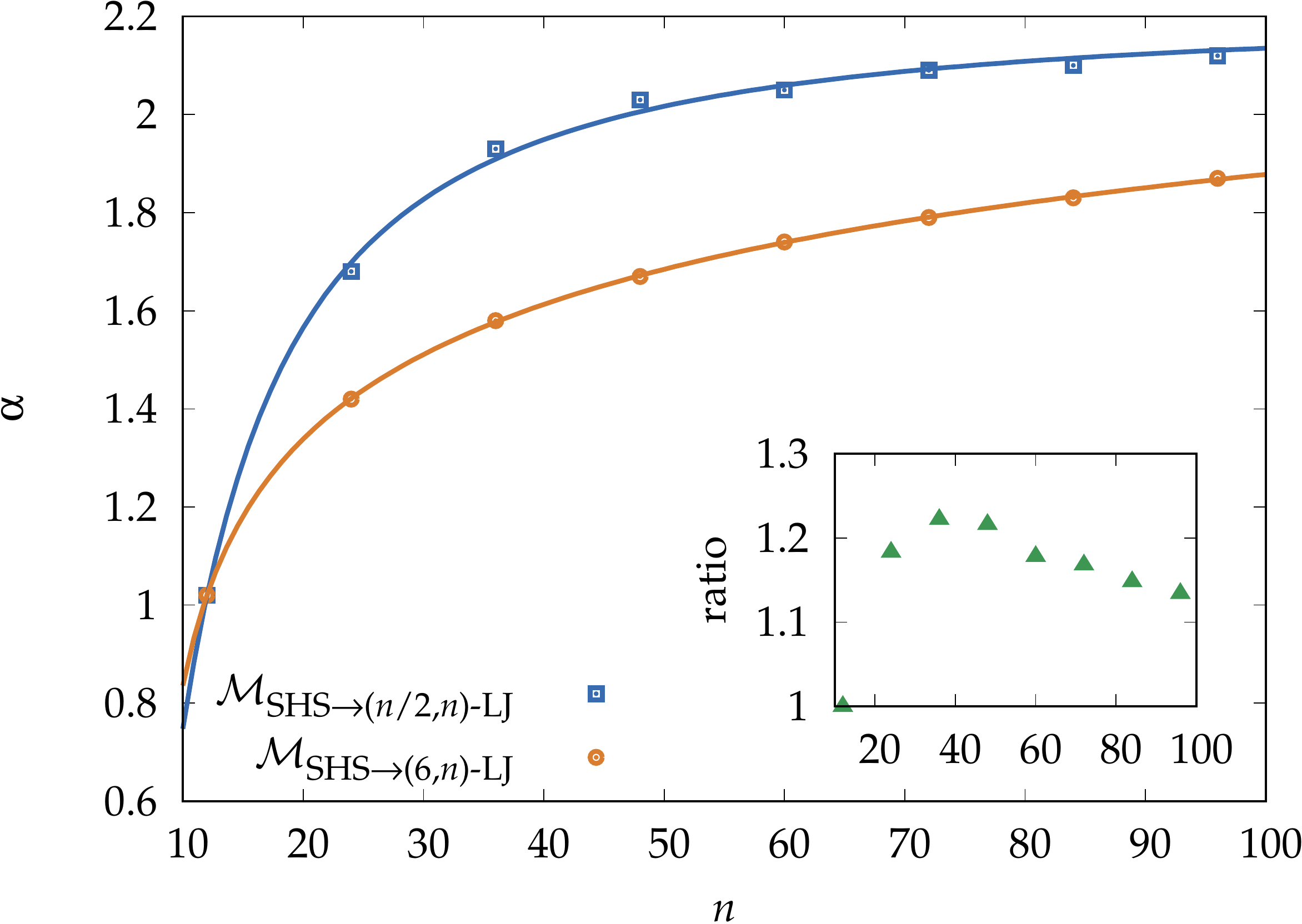}
	\caption{Convergence behaviour of the asymptotic exponential rise rate parameter
	$\alpha$ (Eq.(\ref{eq:Stil})) towards the SHS limit with respect to the LJ exponent $n$. The inlet shows the ratio of
	the two quantities $\alpha(|\mathcal{M}_{\text{SHS}\to (n/2,n)-\text{LJ}}(N)|) /
	\alpha(|\mathcal{M}_{\text{SHS}\to (6,n)-\text{LJ}}(N)|)$.}
    \label{fig:repulsive13-14}
\end{figure}

Using the results for $N \geq 13$ 
from Figure~\ref{fig:expinfty}, we can calculate how $\alpha$ depends on the LJ range parameter $n$.
As shown in Figure~\ref{fig:repulsive13-14}, a general function of the form
\begin{align}
\label{expgrowth}
    \alpha(n)=\alpha_\text{max}+\frac{a}{(n-n_0)^{p}}
\end{align}
fits the results nicely, allowing the prediction of growth behaviour for different
LJ potentials. For $|\mathcal{M}_{(n/2,n)-\text{LJ}}|$, $\alpha_\text{max}$ is
equivalent to $\alpha_\text{SHS}=2.207$. The other adjusted parameters are
$a=-66.588$, $n_0=-3.386$ and $p=1.473$ (Figure~\ref{fig:repulsive13-14}).
We also show the ratio $\alpha(|\mathcal{M}_{\text{SHS}\to (n/2,n)-\text{LJ}}|) /
	\alpha(|\mathcal{M}_{\text{SHS}\to (6,n)-\text{LJ}}|)$ between the two 
	different LJ asymptotic exponential rise rate parameters, which shows that larger
	cluster sizes need to be studied to correctly describe the asymptotic limit.

\begin{figure}
    \centering
    \includegraphics[width=\columnwidth]{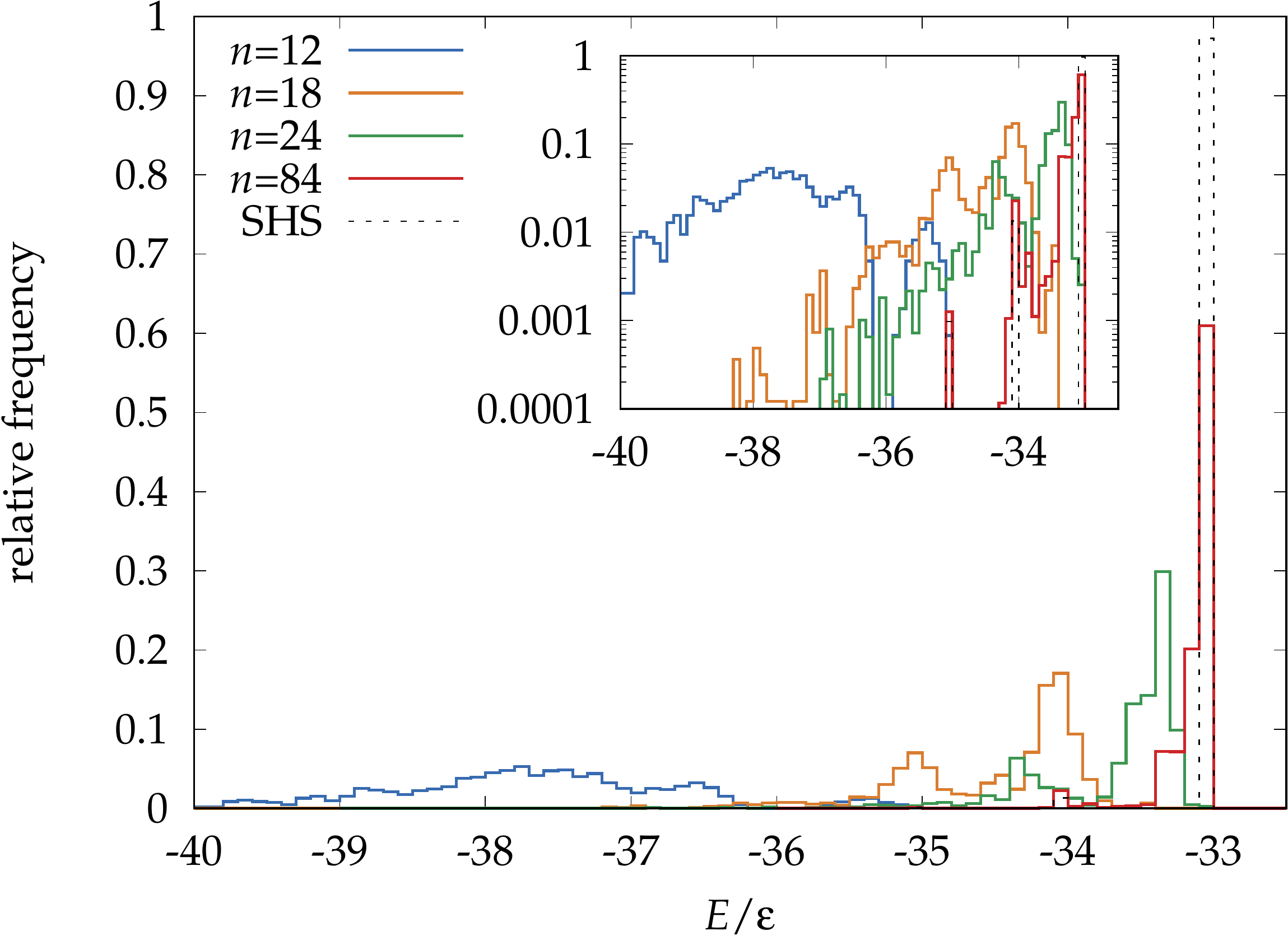}
    \caption{Histogram of the energies (bin size $\Delta\varepsilon=0.1$) of
    minima $\mathcal{M}_{\text{SHS}\to (n/2,n)-\text{LJ}}(N)$ for $N=13$ and
    different exponents $n$ up to the SHS limit. For better visibility, the
    height of the bars are set to $\Delta|\mathcal{M}|/|\mathcal{M}|$ in the
    interval $\Delta(E/\epsilon)$. The inlet shows the same data in logarithmic
    scale.}
    \label{fig:N13-steps}
\end{figure}

The distribution of minima as a function of (free) energy was suggested to be
Gaussian \cite{Sciortino-1999}.  Figure~\ref{fig:N13-steps} shows the energy
distribution of minima for different LJ $(n/2,n)$ potentials derived from SHS
initial structures. We do not see a Gaussian type of distribution; this 
result does not change if we take the free energy at finite temperatures. 
The results indicate a ``phase transition'' in the potential energy landscape away from low-energy to
high energy minima as $n$ increases.
The transition occurs at fairly small $n$. 
Results for the $(9,18)$-LJ potential indicate two HCR-SCA-like maxima that are not present for the $(6,12)$-LJ potential; these are associated with the $N_c = 34$ and $N_c = 35$ SHS clusters,
respectively.
It is also clear that (as expected) the distributions narrow with increasing $n$.

\begin{table}
    \caption{Number of distinct local minima $|\mathcal{M}_\mathrm{SHS}|$ for
    cluster size $N$ (from Refs. \cite{holmes16,hoy_structure_2012,hoy15}) 
    and contact number $N_c$ from the exact enumeration, 
    compared to the number of different structures obtained from a
    geometry optimisation starting from the set $\mathcal{M}_{\mathrm{SHS\to
    LJ}}(N,N_c)$ for a (6,12)-LJ potential. The overall number of unique minima 
 $|\mathcal{M}_\mathrm{SHS\to LJ}|  = \sum_{N_c} |\mathcal{M}_\mathrm{SHS\to LJ} (N_c)| - (\# \rm{\ of\ duplicate\ structures})$    
 is shown in the following column. 
    This result can be compared to the number of unique minima found using the basin-hopping method
    ($|\mathcal{M}_\mathrm{LJ}|$). The difference $\Delta\mathcal{M}=|\mathcal{M}_\mathrm{LJ}| - |\mathcal{M}_\mathrm{SHS\to LJ}|$ 
    is also listed.}
    \label{tab:comp}
    \begin{ruledtabular}
    \begin{tabular}{clccccc}
        $N$ & $N_c$ & $|\mathcal{M}_\mathrm{SHS} (N_c)|$ & $|\mathcal{M}_{\mathrm{SHS\to LJ}}(N_c)|$ & $|\mathcal{M}_\mathrm{SHS\to LJ}|$ & $|\mathcal{M}_\mathrm{LJ}|$     &  $\Delta\mathcal{M}$ \\ \hline 
8  & 18  & 13     				& 8    & 8                     & 8                     & 0                   \\  \\
9  & 21  & 52     				& 20   & 20                    & 21                    & 1                   \\  \\
10 & 23  & 1      				& 1    &                       &                       &                     \\
   & 24  & 259    				& 60   & 62                    & 64                    & 2                   \\
   & 25  & 3      				& 3    &                       &                       &                     \\  \\
11 & 25  & 2      				& 2    & \multirow{6}{*}{165}  & \multirow{6}{*}{170}  & \multirow{6}{*}{5}  \\
   & 26  & 18     				& 6    &                       &                       &                     \\
   & 27  & 1620\footnotemark[1]	& 158  &                       &                       &                     \\
   & 28  & 20     				& 12   &                       &                       &                     \\
   & 29  & 1      				& 1    &                       &                       &                     \\  \\
12 & 28  & 11     				& 6    & \multirow{6}{*}{504}  & \multirow{6}{*}{515}  & \multirow{6}{*}{11} \\
   & 29  & 148    				& 24   &                       &                       &                     \\
   & 30  & 11638  				& 483  &                       &                       &                     \\
   & 31  & 174    				& 69   &                       &                       &                     \\
   & 32  & 8      				& 6    &                       &                       &                     \\
   & 33  & 1      				& 1    &                       &                       &                     \\  \\
13 & 31  & 87     				& 23   & \multirow{8}{*}{1476} & \multirow{8}{*}{1510} & \multirow{8}{*}{34} \\
   & 32  & 1221   				& 100  &                       &                       &                     \\
   & 33  & 95810\footnotemark[1]& 1418 &                       &                       &                     \\
   & 34  & 1318\footnotemark[1] & 293  &                       &                       &                     \\
   & 35  & 96     				& 49   &                       &                       &                     \\
   & 36  & 8      				& 6    &                       &                       &                     \\  \\
14 & 33  & 1      				& 1    & \multirow{8}{*}{4093} & \multirow{8}{*}{(4187)\footnotemark[2]}    & \multirow{8}{*}{(94)\footnotemark[2]}  \\
   & 34  & 707    				& 101  &                       &                       &                     \\
   & 35  & 10537  				& 410  &                       &                       &                     \\
   & 36  & 872992 				& 3939 &                       &                       &                     \\
   & 37  & 10280  				& 1002 &                       &                       &                     \\
   & 38  & 878    				& 237  &                       &                       &                     \\
   & 39  & 79     				& 42   &                       &                       &                     \\
   & 40  & 4      				& 3    &                       &                       &                     \\
    \end{tabular}
    \end{ruledtabular}
    \footnotetext[1]{The largest value for $|\mathcal{M}_\mathrm{SHS}|$ has been taken from 
    Refs.~\cite{holmes16,hoy_structure_2012,hoy15}.}
    \footnotetext[2]{Estimated.}
\end{table}%

It is well known that the global minimum for rare gas clusters with 13 atoms is
the ideal Mackay icosahedron \cite{HoareP75,Hoare-1976,hoare79}. Simple
geometric considerations imply that such a symmetric cluster is not possible
for sticky hard spheres; all vertices of a regular icosahedron with unit edge length
lie on a circumscribing sphere with radius $r_c\approx 0.951$, making it
impossible to insert a sphere of the same radius into the center of the
polyhedron.  Therefore, there must be well-defined LJ exponents $(m,n)$ at
which the icosahedral $N = 13$ LJ cluster breaks symmetry to form a rigid cluster.  
For the $n = 2m$ case considered above, this symmetry-breaking occurs at $m \simeq 15$.

We also explored a more realistic extended LJ potential (Eq.\ \ref{eq:ELJ};  Figure~\ref{fig:LJ})
for one of the rare gas dimers (xenon) 
in comparison with other LJ potentials. We see that the repulsive part agrees
nicely with the conventional (6,12)-LJ potential, while for $r > 1$
the extended LJ potential is slightly less attractive. 
This change should lead to an increase in the number of local minima compared to the conventional
(6,12)-LJ potential. We find that this is indeed the case, i.e.
$|\mathcal{M}_\mathrm{SHS\to ELJ}|=\{8,21,74,205,685,2179,6863\}$ for
$N=\{8,9,10,11,12,13,14\}$.  For $N=13$ the number of distinct minima is 44\% larger than it is for the simple (6,12)-LJ potential,
which shows that $|\mathcal{M}(N)|$ is rather sensitive to the potential chosen.
Hence, to correctly describe the topology of real systems, one has to take care
of the correct form of the 2-body contribution (as well as higher $n$-body
contributions) \cite{Schwerdtfeger-2016}.

\subsection{(6,12)-Lennard-Jones clusters from basin-hopping} 

Table~\ref{tab:comp} shows the number of distinct minima found by our cluster
geometry optimisation procedure using the (6,12)-LJ potential compared to
results from exact enumeration for SHSs and from basin-hopping for the (6,12)-LJ
potential.  As the SHS clusters for a specific $N$ value can be grouped by
their contact number $N_c$, the geometry optimisations were carried out
separately for each group of $\mathcal{M}_\mathrm{SHS}(N_c)$. Hoy
\cite{hoy_structure_2012,hoy15} and Holmes-Cerfon~\cite{holmes16}  reported
slightly different results
for $N=11$ and $N=13$; we find that upon geometry optimisation,
their datasets yield the same final clusters
$|\mathcal{M}_{\mathrm{SHS\to LJ}}(N_c)|$.  As identical LJ clusters appear in
multiple groups with different contact numbers, we remove the duplicates to
create the set $\mathcal{M}_\mathrm{SHS\to LJ}$ of distinct minima, which can
be directly compared to the set of LJ minima $\mathcal{M}_\mathrm{LJ}$ obtained
from the basin-hopping method. It should be noted that including the hypostatic
clusters and the different $|\mathcal{M}_\mathrm{SHS}|$ for $N=11$ and $N=13$
from Ref.~\cite{holmes16} did not change our results, implying that hypostatic
clusters are not an important feature for the LJ energy landscape.

Interestingly, our gradient-based minimisation procedure starting from the SHS
packings does not in general lead to a complete set of LJ minima; the mapping
from SHS minima to LJ minima is non-injective and non-surjective.  Clearly,
some structural motifs found in LJ clusters are not found in SHS clusters and
vice versa, and the topology of the hypersurface changes in a non-trivial
fashion from SHS to LJ.  However, it is surprising that the fraction of
structures that are missed by this optimisation procedure is so small (see
Table~\ref{tab:seeds}). To gain further insight, we analysed the energetics and
structure of the unmatched clusters in more detail.

\begin{table}
    \caption{Range $[E_0,E_\text{max}]$ of the energy spectrum of all LJ
    minima, position of the second lowest minimum structure $E_1$ and position
    of the first unmatched (UM) structure $E_0^\text{UM}$ relative to the
    respective global minimum (in reduced units and $E_0=0$).}
    \label{tab:energies}
    \begin{ruledtabular}
        \begin{tabular}{lccc}
        $N$ & $E_\text{max}$ & $E_1$ & $E_0^\text{UM}$ \\\hline
        8   & 1.04   & 0.06    & -           \\
        9   & 2.08   & 0.84    & 1.19        \\
        10  & 3.13   & 0.87    & 2.22        \\
        11  & 4.22   & 0.85    & 2.27        \\
        12  & 6.16   & 1.62    & 3.38        \\
        13  & 9.26   & 2.85    & 6.14        \\
        \end{tabular}
    \end{ruledtabular}
\end{table}

\begin{table}
    \caption{Number of missing structures after optimisation belonging to the
    same "seed" (Fig.\ \ref{fig:seeds}). $N=8$ is excluded because all LJ minima were
    found starting from the SHS model.}
    \label{tab:seeds}
    \begin{ruledtabular}
    \begin{tabular}{llllll}
        seed      & $N=9$   & $N=10$  & $N=11$  & $N=12$  & $N=13$  \\ \hline
        a         & 1    & 1    & -    & 3    & 8    \\
        b         & -    & 1    & 3    & 4    & 12\footnotemark[1]   \\
        c         & -    & -    & 1    & 1\footnotemark[1]    & -    \\
        d         & -    & -    & 1    & 1    & 5    \\
        e         & -    & -    & -    & 1    & 6    \\
        f         & -    & -    & -    & 1    & 1    \\
        remaining & -    & -    & -    & -    & 2    \\ 
        total     & 1    & 2    & 5    & 11   & 34   \\
        \%        & 4.76 & 3.13 & 2.94 & 2.14 & 2.25 \\ 
    \end{tabular}
    \end{ruledtabular}
        \footnotetext[1]{Some structures do not resemble a perfect capped
        cluster, but undergo a slight rearrangement. Specifically, two structures belonging to seed (b) and one structure belonging to seed (c) were found to deviate slightly from the perfect arrangement, but minor rearrangements of these structures lead to the desired geometry and they can be reasonably associated with these seeds.}
\end{table}%

\begin{figure*}
    \centering
    \subfloat[$N=11$]{\includegraphics[width=.32\textwidth]{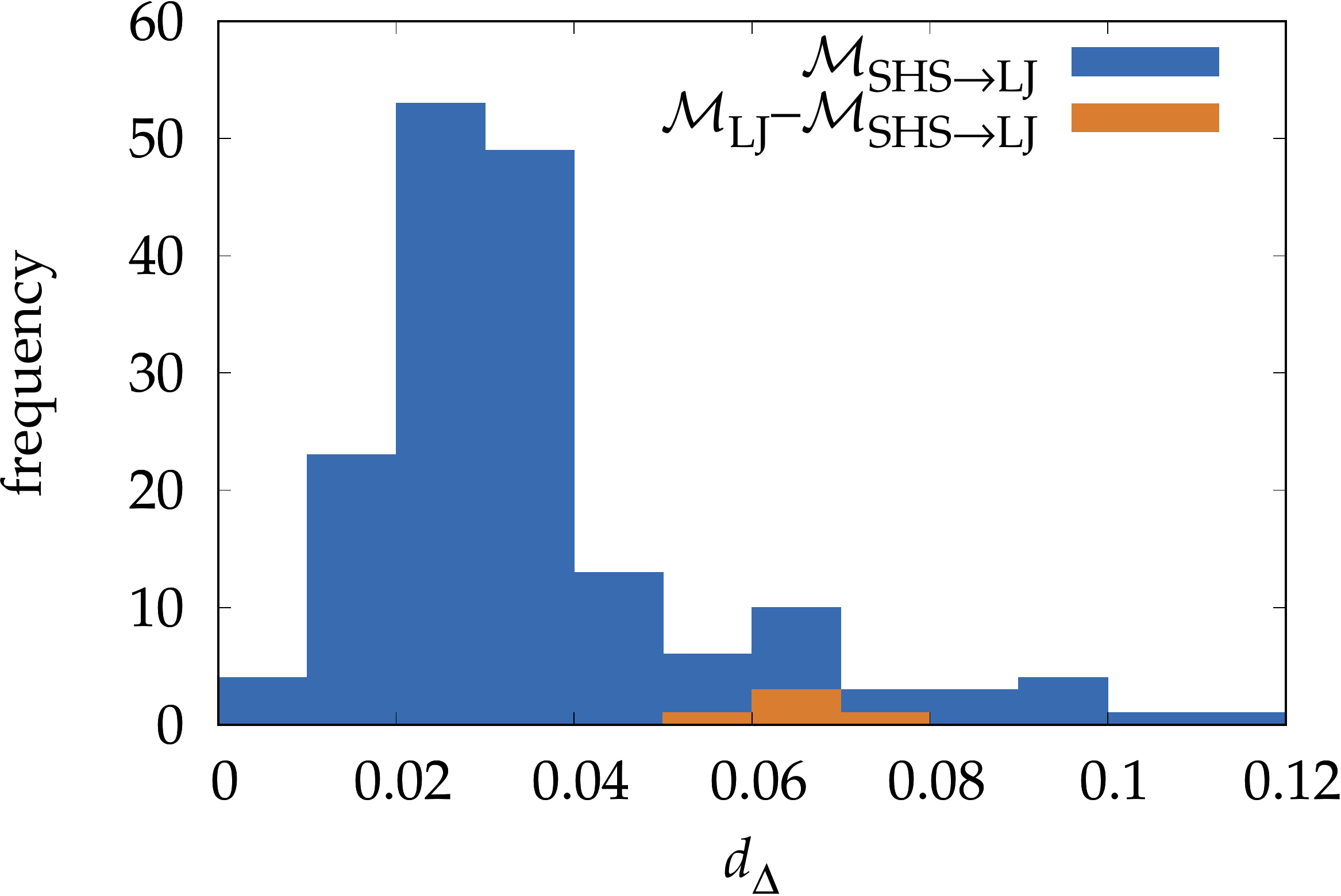}}
    \subfloat[$N=12$]{\includegraphics[width=.32\textwidth]{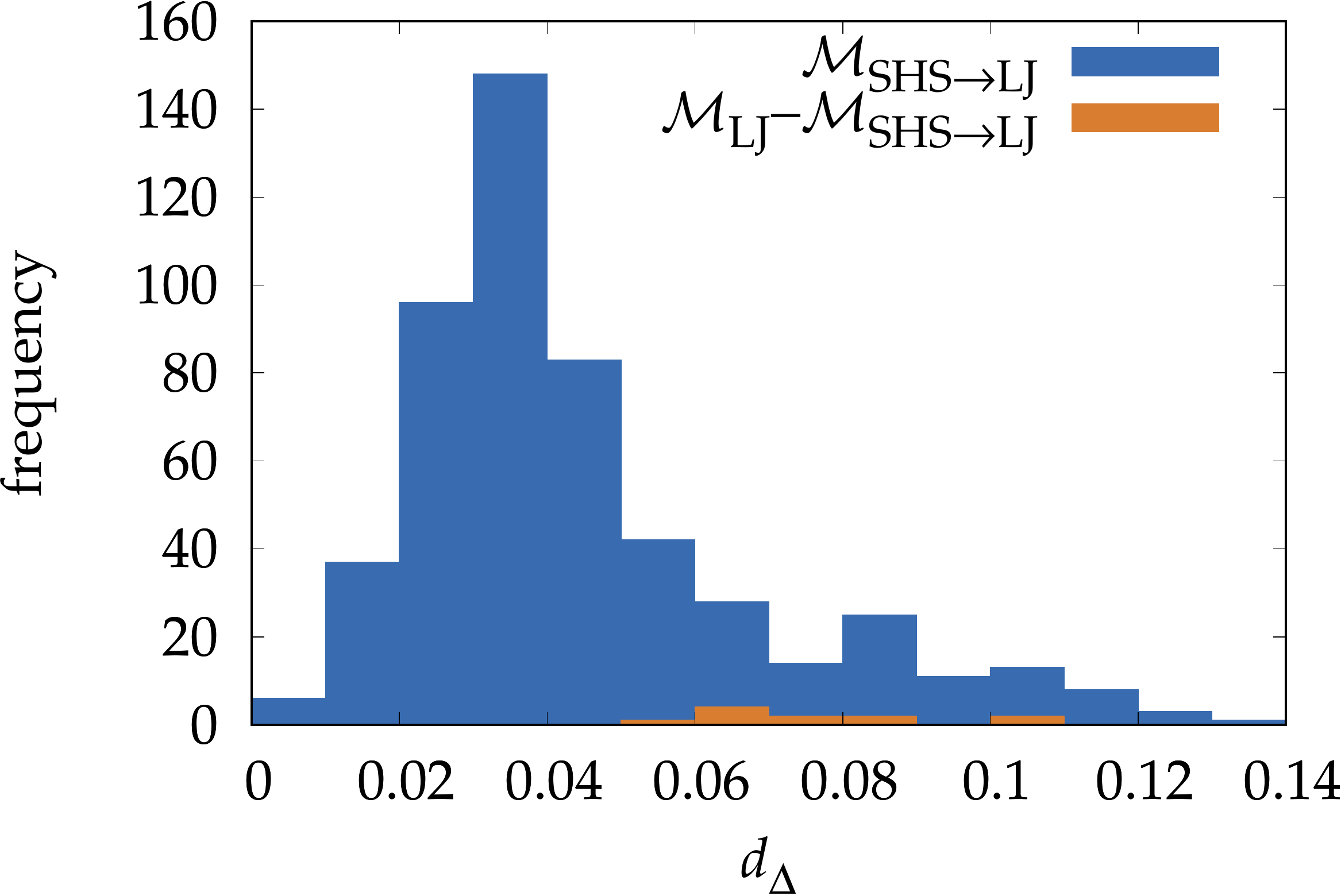}}
    \subfloat[$N=13$]{\includegraphics[width=.32\textwidth]{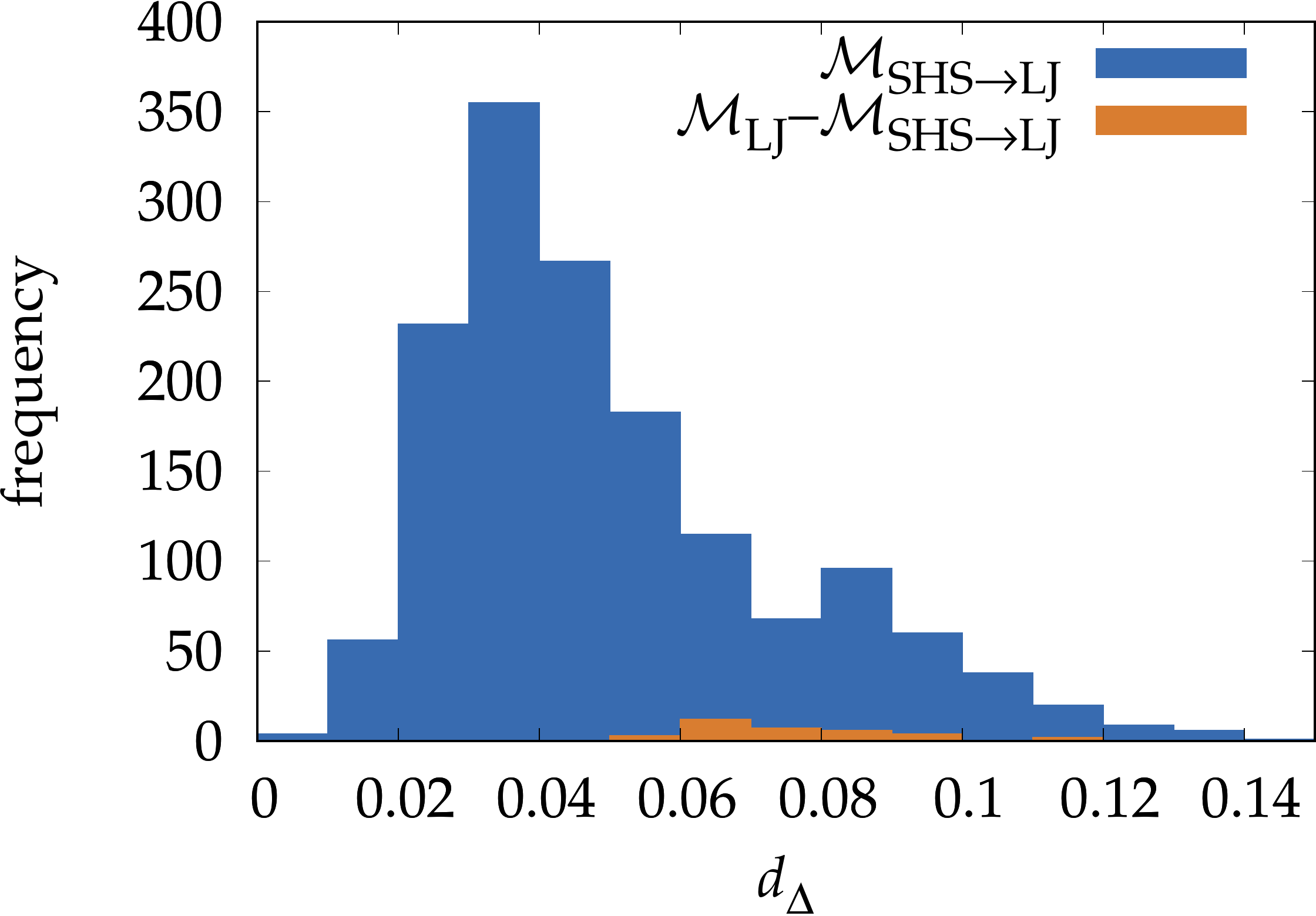}}
    \caption{Histograms of the difference between the longest and shortest bond
    distances $d_\Delta=d_\text{max}-d_\text{min}$ for the complete set of
    distinct LJ minima $\mathcal{M}_\text{LJ}(N)$ for $N=\{11,12,13\}$. Orange
    bars give the number of distinct structures not contained in
    $\mathcal{M}_\mathrm{LJ}$ as obtained from the basin-hopping algorithm.}
    \label{fig:bondlength-variance}
\end{figure*}%

Figure~\ref{fig:bondlength-variance} shows an analysis of the difference
between the longest to the shortest bond lengths $d_\Delta=d_{\rm max}-d_{\rm
min}$ obtained for the largest clusters in $\mathcal{M}_{\mathrm{LJ}}$ with
$N=\{11,12,13\}$ \footnote{We define spheres that have a equilibrium distance
between $0.9-1.1$ to be bound.}.  The histograms show that the clusters most
commonly have a $d_\Delta$ of about $0.03$.  In contrast, as shown by the
orange bars, the unmatched structures have significantly larger $d_\Delta$
values of at least $0.05$, with most of them having $d_{\Delta} \simeq 0.06$.
This is a first indication of why these structures are not found by starting
from SHS packings. The latter only form bonds of length one, and a large
variation in bond length could imply that a SHS packing similar to the LJ
structure does not exist as the SHS boundary conditions are not satisfied.  The
data in Table~\ref{tab:energies} show that the unmatched (UM) structures for a
specific $N$ value have much higher energies compared to the one of the global
minimum (which is set to zero, i.e.  $E_0=0$).  They are always positioned in
the upper half of the energy spectrum, making them energetically unfavorable. 
However, we
could not find any correlation between $d_\Delta$ and the energetic
position of the LJ clusters.

\begin{figure}
    \centering
    \subfloat[]{\includegraphics[width=0.33\columnwidth]{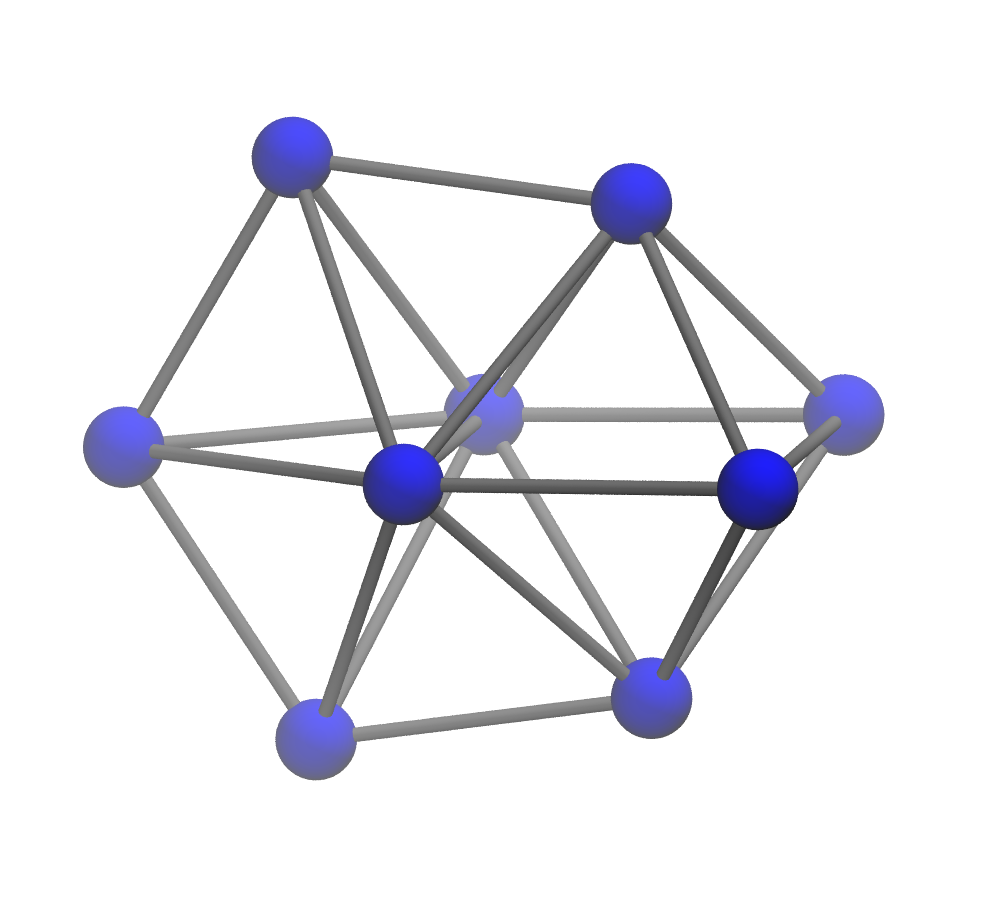}}
    \subfloat[]{\includegraphics[width=0.33\columnwidth]{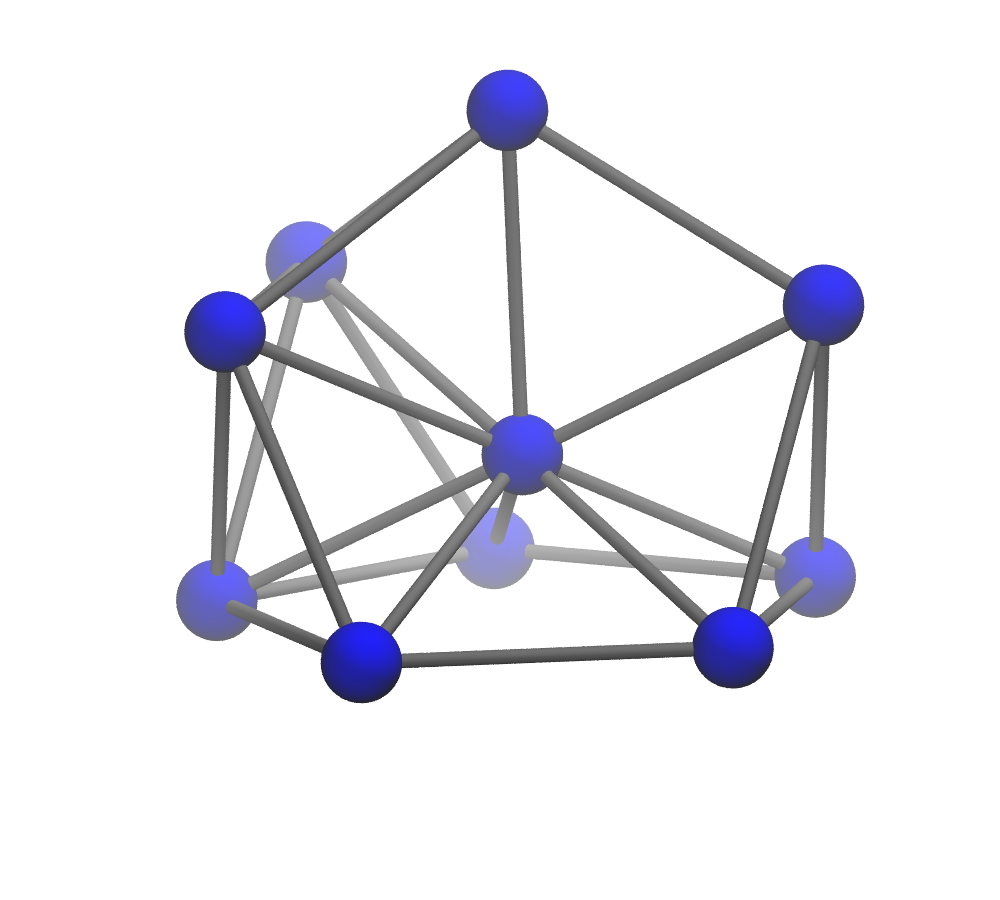}}
    \subfloat[]{\includegraphics[width=0.33\columnwidth]{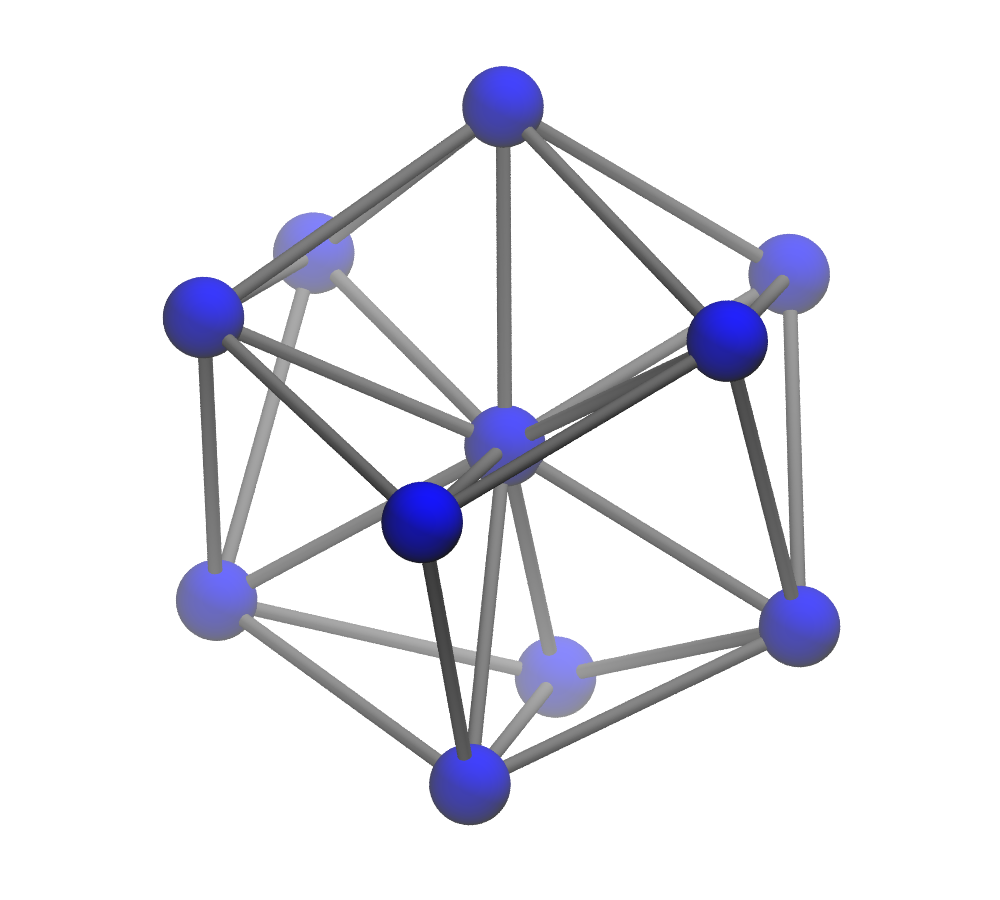}}\\
    \subfloat[]{\includegraphics[width=0.33\columnwidth]{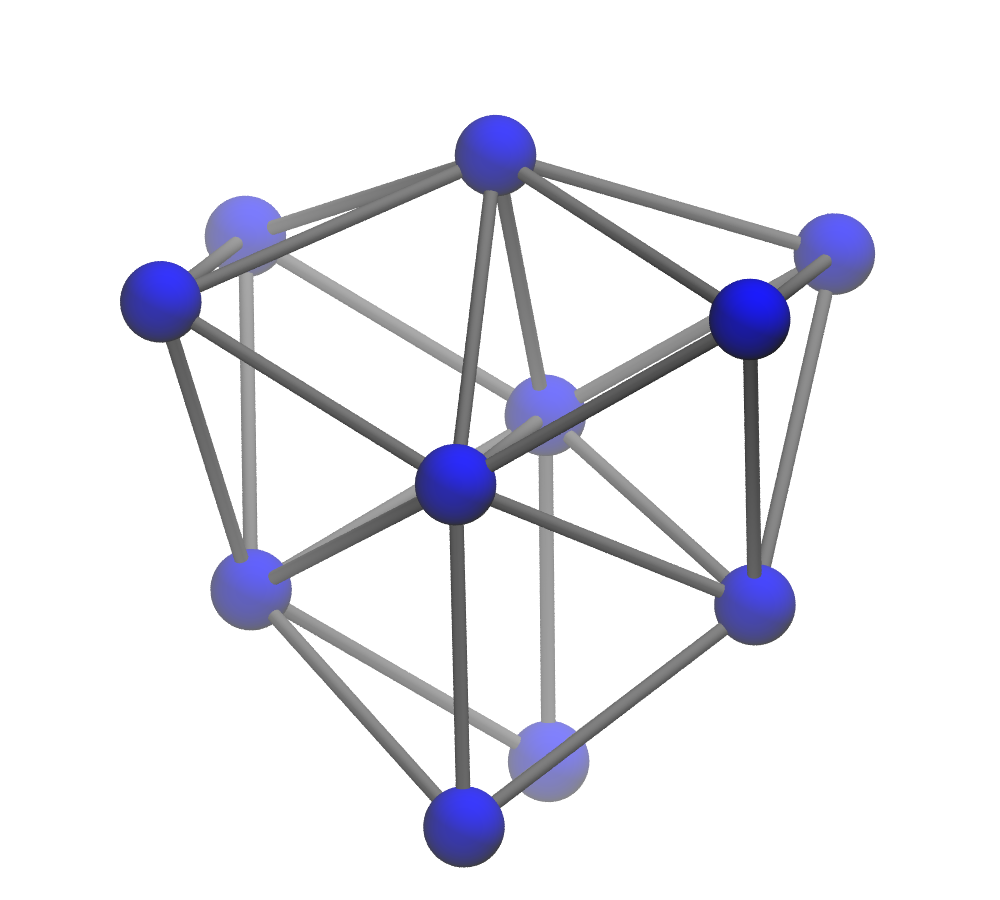}}
    \subfloat[]{\includegraphics[width=0.33\columnwidth]{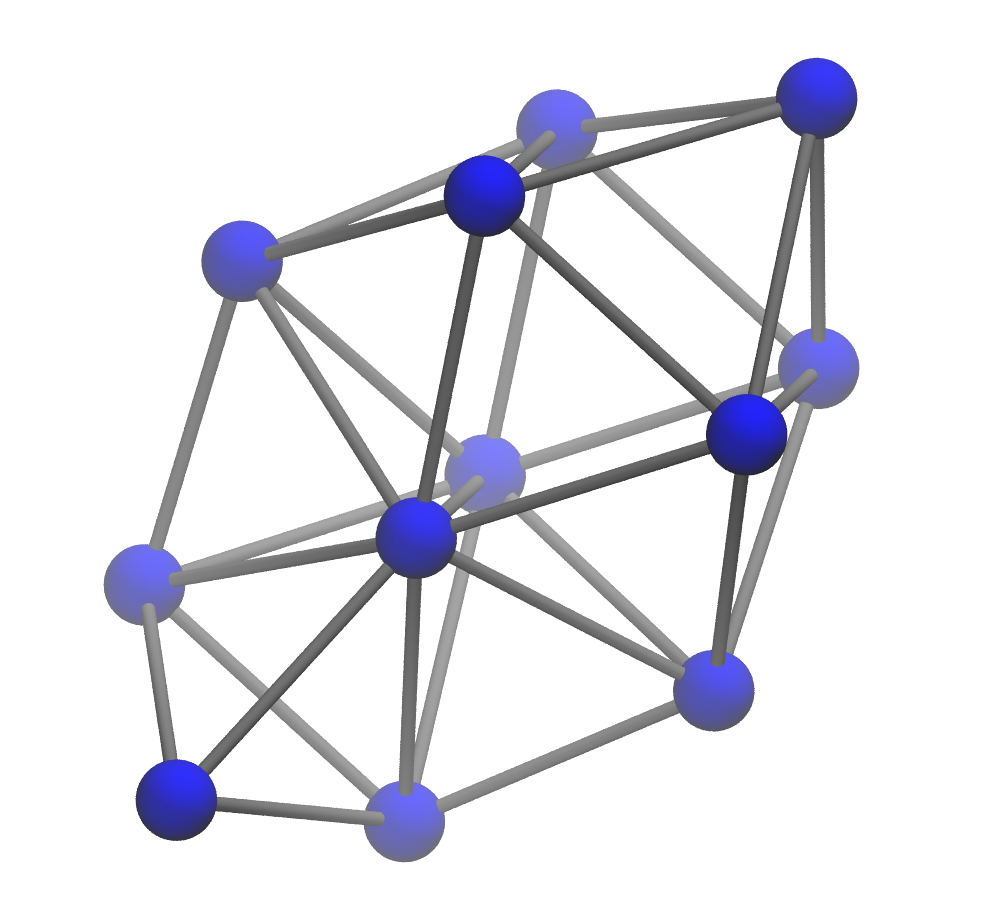}}
    \subfloat[]{\includegraphics[width=0.33\columnwidth]{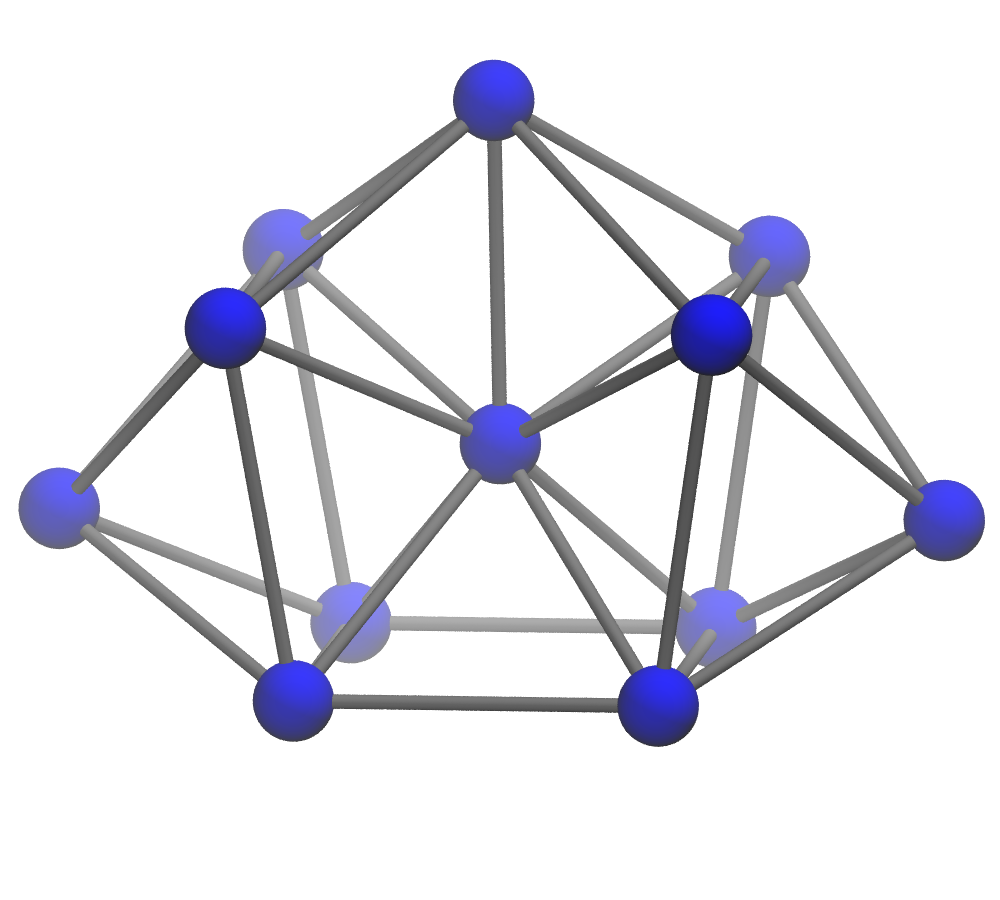}}
    \caption{Graphical representations of the structures that are starting new
    seeds, but are not contained in $\mathcal{M}_\mathrm{SHS\to LJ}$. See
    Table~\ref{tab:seeds} and text for more details.}
    \label{fig:seeds}
\end{figure}%

Last, we checked the geometries of the missing structures in more detail.
As it turns out, almost all of the missing stable LJ clusters can be created
from a smaller set of missing clusters by capping some of their triangular
faces. Therefore, these groups of clusters can be referred to as ``seeds''
\cite{arkus_deriving_2011}. The corresponding starting structures of each seed
are shown in Figure~\ref{fig:seeds}. 
None of these structures are stable SHS packings. For
example, structure (d) can be described as three octahedra connected via
triangular faces sharing one edge. Geometric considerations \cite{arkus_deriving_2011,hoy_structure_2012} immediately show
that this structure cannot be a stable SHS packing;
the dihedral angle in an octahedron is approximately $109.5^\circ$, which means three
octahedra only fill $328.5^\circ$ of a full circle, leaving a gap between two
faces.

Table~\ref{tab:seeds} shows the number of missing minima belonging to each seed.
Over 60~\% of the unmatched structures belong to seeds (a) and (b).  
From a graph theoretical point of view \cite{arkus_minimal_2009,arkus_deriving_2011},
grouping structures into seeds means that all structures belonging to the same
seed contain the graph of the starting structures as a subgraph in their
respective connectivity matrix.  This approach simplifies the analysis to a
great extent, as the feature that prevents the structures from being found by
geometry optimisation is the same for each of the structures arising from a
specific seed.  
The smallest unmatched structures that cannot be associated with any of seeds (a)-(f) have $N=13$;
these could be the starting structures for two new seeds.

Finally, we note that the starting SHS minima in our optimisation procedure are
not stationary points on the LJ hypersurface, and we therefore optimise to most
but not all local and available LJ minima. This observation explains why some high-energy
structures were not found by our optimisation procedure. For a smooth change in the topology of the potential energy surface
from SHS to LJ type clusters one has to continuously vary the exponents $(n,m)$
in real space, which is computationally too demanding.


\subsection{The special case of a Gregory-Newton cluster}

We call a cluster ``Gregory-Newton'' (GN) when it belongs to the set of all clusters consisting of 12
spheres kissing a central sphere. The canonical Gregory-Newton cluster is the icosahedron, which is 
perhaps the most common form studied in cluster chemistry and physics \cite{Mackay-1962,Uppenbrink-1991,Martin-1996,conway-2013book}.
We therefore investigate this cluster type in more detail here. 

For monodisperse SHS clusters, the Gregory-Newton argument (as proved by Sch\"utte and van der
Waerden \cite{Schutte-1952}) that no more than 12 equally sized spheres can touch a central sphere of same size holds.
We note that the problem of the number of kissing spheres in $k$ dimensions, or even in three dimensions 
with sphere size smaller than that of the central sphere, remains largely unsolved \cite{conway-2013book,Musin-2012}. 
For unequally sized spheres, some simple results related to spherical codes \cite{phillips12} are  
known; for example, 13 hard spheres of radius $r_s$ can touch a central sphere of unit radius only if $r_s \leq 0.9165$ \cite{phillips12}.
For particles interacting via finite-ranged potentials such as $V^{LJ}_{mn}(r)$, however, the situation is far more complicated since systems minimize energy rather than differences in the distances between neighboring particles, and few general results are known.
Nonetheless, this latter problem is important for understanding real systems such as coordination compounds \cite{Hermann-2007}, which have recently been shown to possess coordination numbers as high as 17 \cite{Kaltsoyannis-2017} or even 20 \cite{Suresh-2016}.

Motivated by these recent results,
we investigated longer range potentials  by decreasing the LJ exponents ($m,n$), to see whether the restriction of no more than 12 kissing equal-sized spheres still holds.
As it is impossible to distribute 13 points on a sphere evenly (there is no triangulation of a sphere
with 13 vertices of degree 5 and 6 \cite{Schwerdtfeger-2017}), we used the Fibonacci sphere
algorithm \cite{gonzalez_measurement_2010,keinert_spherical_2015} to find an
approximate distribution of points on a sphere and added a center sphere. By
optimising the coordinates for this $N=14$ cluster with different LJ exponents
and calculating the distance of every sphere to the center sphere, we can
deduce at which ``softness'' a 13th sphere is (perhaps) allowed to enter the
first coordination shell, i.e.~to touch the center sphere.

\begin{figure}
    \centering
    \includegraphics[width=\columnwidth]{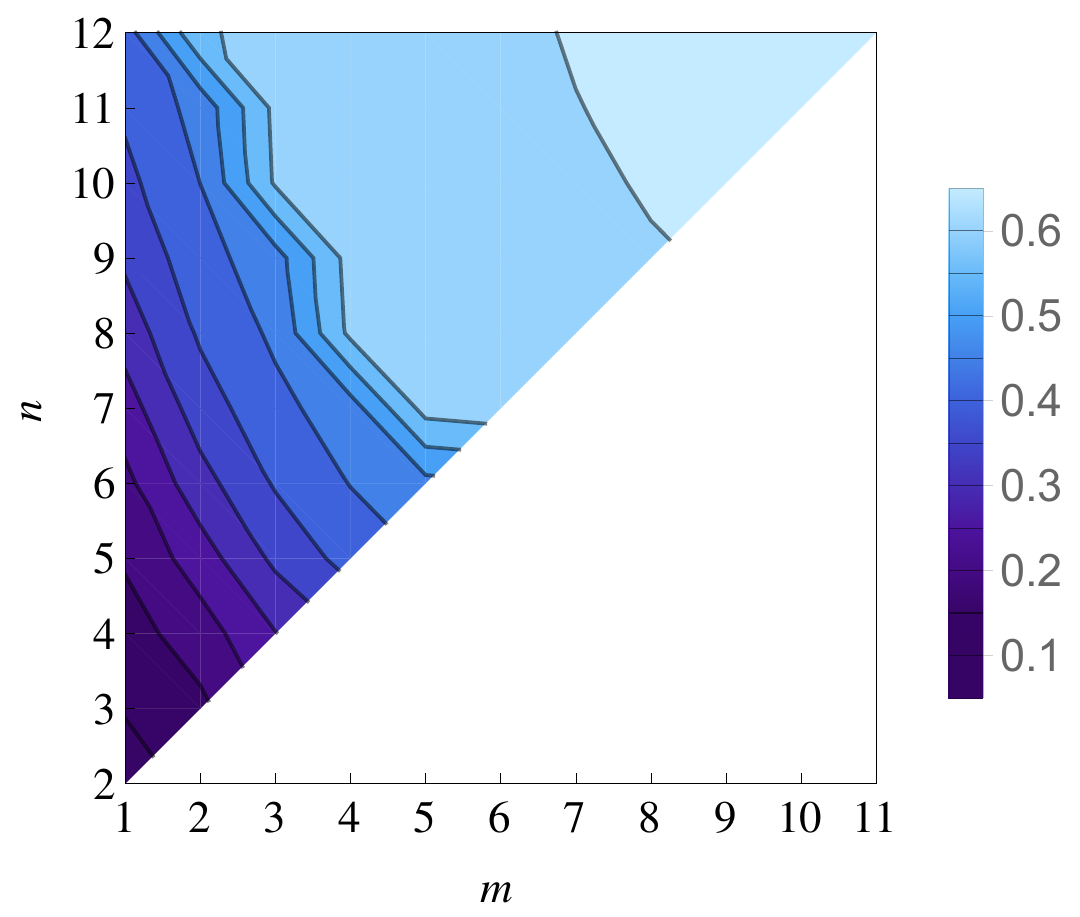}
    \caption{Relation of LJ exponents $m$ and $n$ to the difference of largest
    and smallest center-to-outer sphere (COS) distances. A value of zero would imply that all
    surrounding spheres are touching the center sphere.}
    \label{fig:GregNewt}
\end{figure}

Figure~\ref{fig:GregNewt} shows the difference between the largest and the smallest center-to-outer sphere (COS) distances in relation to the LJ exponents $m$ and $n$. Interestingly, none of the $(m,n)$-LJ
potentials lead to equal distances around a central sphere. While this result
could be due to the lack of symmetry, one sphere is clearly further away from
the central sphere even for the softest ``Kratzer'' (1,2)-LJ potential \cite{Kratzer-1920}. For this potential the
largest and smallest COS distances are $r_{\rm max}=0.882$ and $r_{\rm min}=0.804$,
respectively. While the longest distance only shows up once, the shortest
distance appears twice. All other 10 distances fall in the range between $r=0.845$
and $r=0.861$. The $r_{\rm max} / r_{\rm min}$ ratio is 1.097 and much shorter compared to 
$r_{\rm max} / r_{\rm min}=\sqrt{2}$ for the closed packed lattice, or the shortest
distance possible for the SHS system which is $r_{14}^{\rm GN}=1.347$ (see discussion below).
Hence the 13th sphere ``almost'' touches the center sphere.

Note that all COS distances for the $N=14$ (1,2)-LJ cluster are significantly
shorter than $r=1$, due to the $N(N-1)/2$ attractive 
two-body interactions and the softness of the potential. 
For infinite (e.g. body-centered cubic or close-packed) lattices of particles interacting via $V^\mathrm{LJ}_{mn}(r)$ with $n> m >3$,
one can prove \cite{Schwerdtfeger-2006} that the nearest neighbor distance is
\begin{equation}
    r_\mathrm{NN}(m,n)=\left( L_n L_m^{-1}\right)^\frac{1}{n-m}. 
    \label{eqn:lattice}
\end{equation}
Here $L_n$ is the Lennard-Jones-Ingham lattice coefficient for a specific
lattice determined from 3D lattice sums.  Since $L_n<L_m$ for $n>m$, we see
that $r_\mathrm{NN}<1$, and $\lim\limits_{m,n\rightarrow
\infty}r_\mathrm{NN}(m,n)=1$.  The shortest distances found in (6,12)-LJ
clusters $r_{\rm min}(N)$ are: $r_{\rm min}(8)=0.986767$, $r_{\rm
min}(9)=0.964404$, $r_{\rm min}(10)=0.964382$, $r_{\text min}(11)=0.956345$,
$r_{\rm min}(12)=0.947842$, and $r_{\rm min}(13)=0.952179$.  Surprisingly,
$r_{\rm min}(12)$ is smaller than $r_\mathrm{NN}(6,12)$ for typical crystalline
lattices; $r_\mathrm{NN}(6,12)$ values are $0.95066$, $0.95186$ and $0.97123$
for simple cubic, body-centered cubic and close-packed lattices, respectively.
This result shows that stable clusters do not necessarily have longer bonds
compared to the solid state, where we expect a maximum in interaction energy
per atom.

Finally, we relate the above results back to the motifs present in the HCR-SRA
limit by focusing on $N = 13$ and $N = 14$ SHS clusters from
Ref.~\cite{holmes16}. This set contains all nonisomorphic SHS structures that
can be considered GN clusters ($N=13$) and the $N=14$ structures that can be
derived from them by attaching a 14th sphere. We find a surprisingly large
number ($737$) of nonisomorphic $N = 13$ GN-SHS structures
($\{724,10,1,2\}$ for $N_c=\{33,34,35,36\}$), that all optimise
to the ideal icosahedral arrangement ($I_h$ symmetry) if a (6,12)-LJ potential
is applied. An even larger number of clusters exists for $N=14$ ($14529$),
which is $\approx 0.016|\mathcal{M}_\mathrm{SHS}(14)|$. All of these structures
optimise to just one of two possible (6,12)-LJ minima of GN type. The first is
the Mackay icosahedron capped at one of its triangular faces, and the second is
an elongated pentagonal bipyramid (belonging to the class of Johnson solids)
with the 14th sphere capping a square face.

\begin{figure}
    \centering
    \subfloat[$r_{14}^{\rm GN}=1.34715$, $N_c=39$\label{subfig:short-greg-newton}]{\includegraphics[width=0.5\columnwidth]{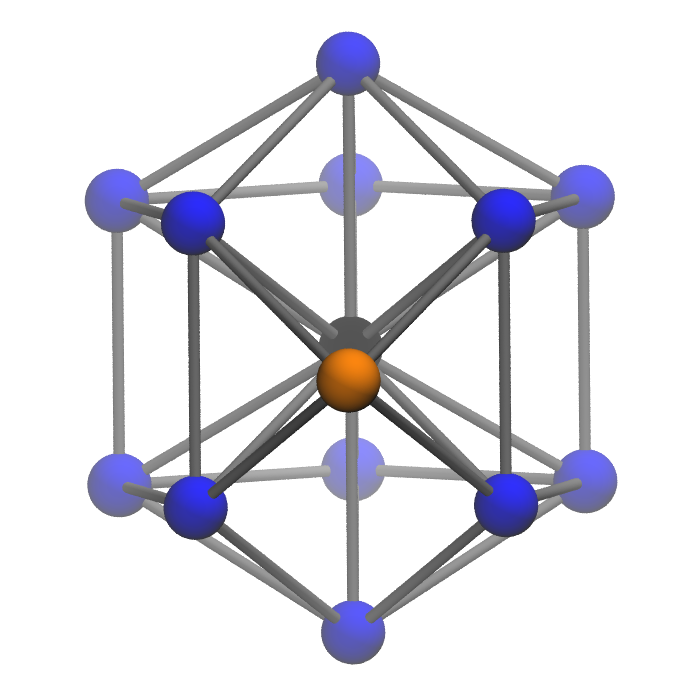}}
    \subfloat[$r_{14}^{\rm GN}=1.37515$, $N_c=36$\label{subfig:2ndshort-greg-newton}]{\includegraphics[width=0.5\columnwidth]{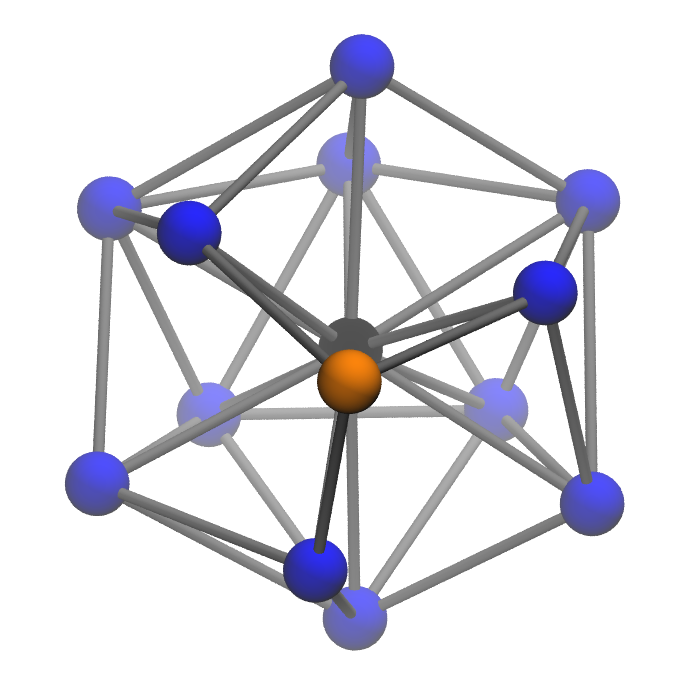}}\\
    \subfloat[$r_{14}^{\rm GN}=\sqrt{2}$, $N_c=40$\label{subfig:sqrt2-greg-newton}]{\includegraphics[width=0.5\columnwidth]{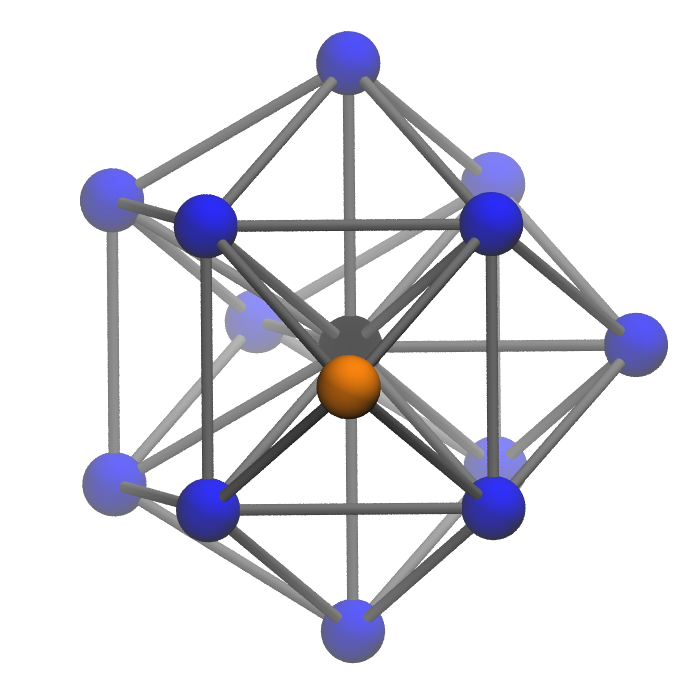}}
    \subfloat[$r_{14}^{\rm GN}=\sqrt{\frac{8}{3}}$, $N_c=39$\label{subfig:sqrt83-greg-newton}]{\includegraphics[width=0.5\columnwidth]{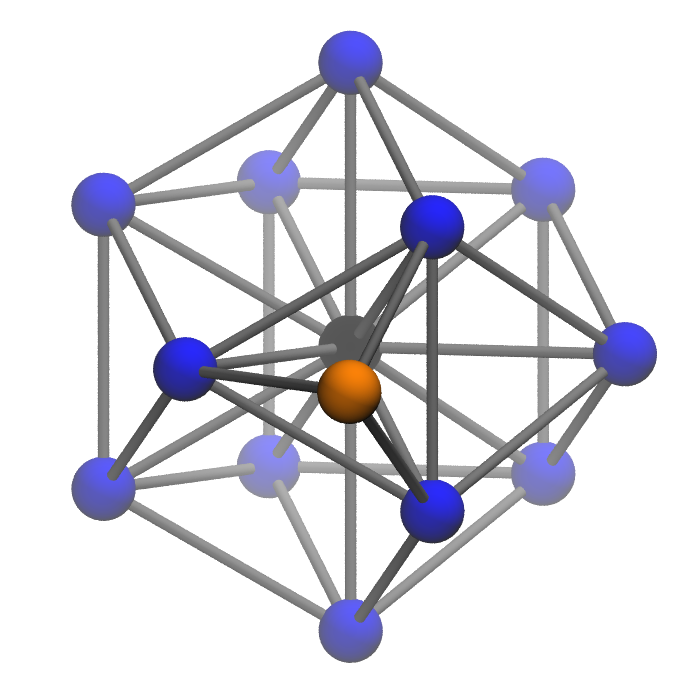}}\\
    \caption{Graphical representations of SHS packings with $N=14$, where a
    center sphere is maximally contacting. The orange sphere in each
    cluster is the 14th outer sphere, not able to touch the center sphere (in black).
    (a) distorted elongated pentagonal bipyramid (Johnson solid); (b) distorted icosahedron; (c) hcp capped on a square; (d) hcp capped on a triangle.}
    \label{fig:N14}
\end{figure}

Most of these $N=14$ clusters are minimally rigid ($N_c=3N-6=36$), while only a
few are hyperstatic ($N_c > 3N-6$) and none are hypostatic ($N_c < 3N-6$).
There are $\{14369,144,8,6,2\}$ such clusters with $N_c=\{36,37,38,39,40\}$ and
$N=14$.  The clusters with $N_c=40$ are hcp and fcc core-shell structures
capped at a square face; these arrangements maximise $N_c$. Most of the clusters
with $N_c=\{38,39\}$ are deformed versions of the elongated pentagonal
bipyramid mentioned above, indicating that this arrangement is a favored route
to these intermediate-energy structures.  However, $N_c=39$ also contains hcp
and fcc structures capped at a triangular face.  The first example of a cluster
derived from a perfect icosahedral symmetry shows up at lower value $N_c=37$
(!).  Representative examples for clusters with high contact numbers are
depicted in Figure~\ref{fig:N14}.  

\begin{figure}
    \centering
    \includegraphics[width=\columnwidth]{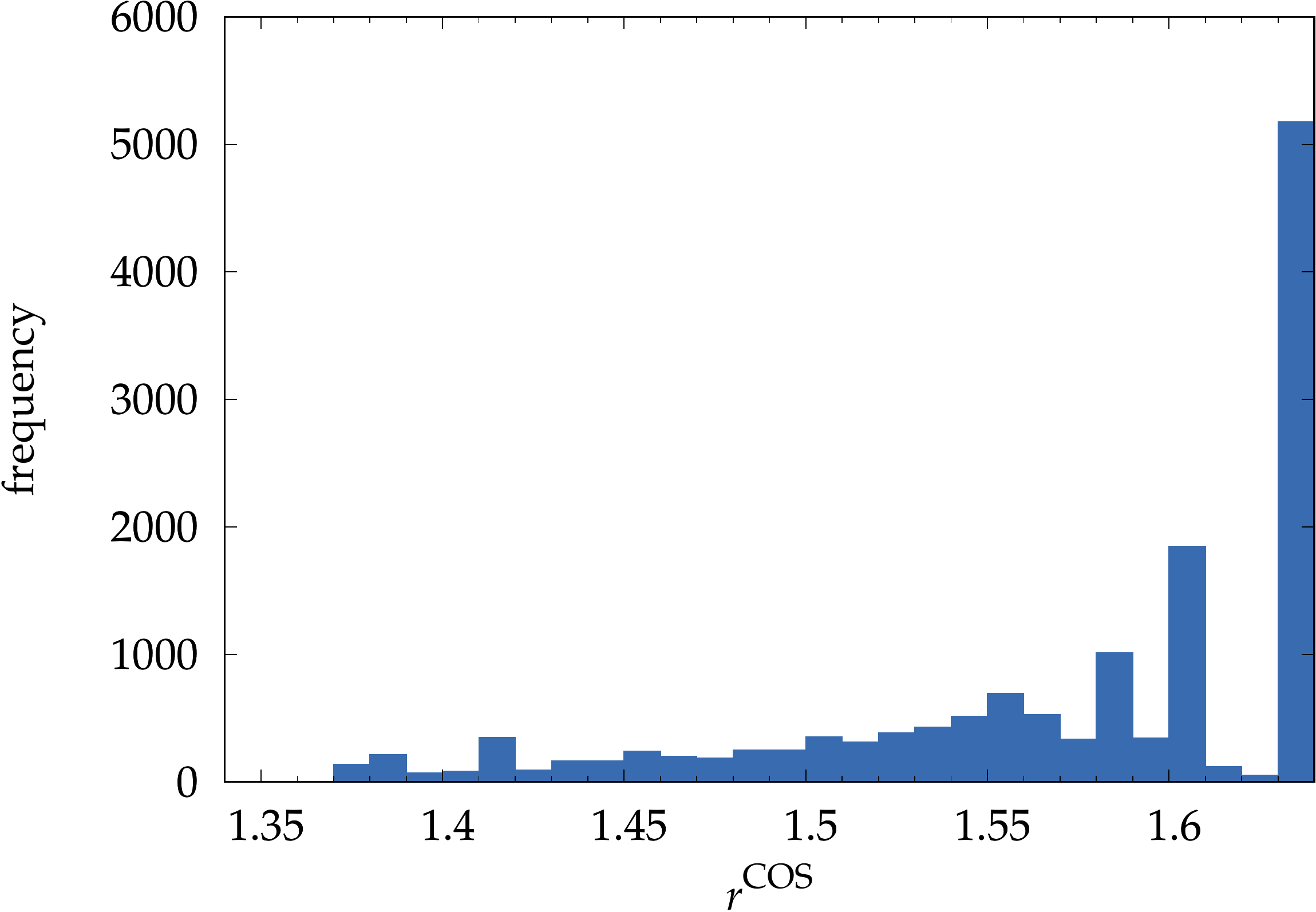}
    \caption{Frequency of distances from the cluster center to the most distant
    sphere for all Gregory-Newton-like clusters contained in the structures
    from Ref.~\cite{holmes16}. The width of the bars is $0.01$.}
    \label{fig:greg-newton}
\end{figure}

Surprisingly, the $N = 14$ cluster with the closest central-to-outer sphere
(COS) distance $r_{\rm min}^{\rm COS}$ was not known. Here we close this gap by
determining the COS distance for all Gregory-Newton type clusters.  
We find one single cluster with $r_{\rm min}^{\rm COS}=1.3471506281091$.
Its structure [Fig.\ \ref{fig:N14}(a)] is similar to the elongated pentagonal bipyramid (a Johnson solid) with one of the square faces stretched to form a regular rectangle.
The 14th sphere caps this deformed face, becoming the vertex of a deformed
octahedron and allowing the outer sphere to get closer to the central sphere.
The next-smallest-$r^{\rm COS}$ cluster ($r^{\rm COS} = 1.37515$) is shown in Fig.~\ref{subfig:2ndshort-greg-newton}.
It does not belong to the category of the clusters derived from the elongated
pentagonal bipyramid, but instead can be described as being icosahedral-like.
The short distance is achieved by attaching the 14th sphere to 3 spheres that
do not form a face of the cluster (because they are separated by a distance
larger than $1$.)

As shown in Figure \ref{fig:greg-newton}, 
the distribution of $r^{\rm COS}$ values for the full set of GN clusters is shown in Figure  \ref{fig:greg-newton}.
Motifs with larger $r^{\rm COS}$ are far more prevalent.
For example, the peak at $r^{\rm COS} = 1.41$ corresponds to
structures where the 14th sphere is touching 4 other spheres that are part of a tetragonal pyramid, therefore forming a regular
octahedron with a tip-to-tip distance of $\sqrt{2}$
(Fig.~\ref{subfig:sqrt2-greg-newton}).  
The maximum $r^{\rm COS}$ value ($1.63$)
corresponds to capping triangular faces, so that the most distant sphere is
part of a regular trigonal bipyramid with a height of $\sqrt{8/3}$
(Fig.~\ref{subfig:sqrt83-greg-newton}).  The structures in the bars at
$1.60,1.58$ and $1.55$ are derived from the regular trigonal bipyramid and
result from breaking its axial bonds. 
In these structures, the more bonds are broken, or the further
the axial spheres are separated, the shorter the center-to-outer sphere distance becomes.

\section{Conclusions}

We have characterized the sets of $(m,n)$-LJ-potential minima obtained using complete sets of nonisomorphic SHS packings with $8 \leq N \leq 14$ \cite{arkus_minimal_2009,arkus_deriving_2011,hoy_structure_2012,hoy15,holmes16} as initial states for energy minimization.
The number of distinct minima (i.e.~excluding
permutation-inversion isomers) is far smaller than the number
of SHS packings for the standard Lennard-Jones exponents $(m,n) = (6,12)$, but approaches the SHS limit from below as ($m,n$) increase.
We characterized how the number of distinct minima $\mathcal{M}(N)$ increases with cluster size $N$ by determining Stillinger's rise rate parameter $\alpha$ (Eq.\ \ref{eq:Stil} \cite{Stillinger-1999}).
The increase of $\alpha$ from $\approx 1.1$ for (6,12)-LJ clusters to $\approx 2.2$ for SHS clusters is described by a simple functional form (Eq.\ \ref{expgrowth}).
All these results  can be understood in terms of a smooth progression of the $(m,n)$-LJ energy landscape towards the SHS energy landscape as $(m,n)$ increase.

Using a more realistic extended LJ potential obtained from coupled cluster
calculations for the xenon dimer \cite{Schwerdtfeger-2006,jerabek_relativistic_2017} leads to $\mathcal{M}$ values close to those obtained for
the (6,12)-Lennard-Jones potential, but our results indicate the the topology of the energy hypersurface is very sensitive to the
model potential applied. 
For softer potentials, we showed that it is still unfavourable for a 13th outer sphere to touch the center
sphere.
Indeed, the Gregory-Newton argument still holds true for even the softest $(m,n) = (1,2)$ potential.

Finally, we compared our optimisation results to the previously published
results for the (6,12)-LJ potential. The mapping from $\mathcal{M}_\text{SHS}$
to $\mathcal{M}_\mathrm{SHS\to LJ}$ is non-injective and non-surjective, however,
the number of structures missed by the optimisation procedure is relatively
small. The unmatched structures belong to the high energy region of
the potential energy hypersurface and possess rather large variations in their
bond lengths. An analysis of their geometries revealed that most of the larger
structures can be constructed from a smaller cluster by capping some
of the triangular faces. This procedure effectively sorts almost all unmatched structures
into six seeds for clusters up to $N=13$.\\

\section{Acknowledgements}
We acknowledge financial support by the Marsden Fund of the Royal Society of New Zealand (MAU1409).
DJW gratefully acknowledges financial support from the EPSRC. PS acknowledges financial support by the Centre for Advanced Study at the Norwegian Academy of Science and Letters (Molecules in Extreme Environments Research Program). We thank Drs. Lukas Wirz and Elke Pahl 
for useful discussions.

\bibliography{KS}
\end{document}